\documentstyle[11pt,aaspp4]{article}
\def\puncspace{\ifmmode\,\else{\ifcat.\C{\if.\C\else\if,\C\else\if?\C\else%
\if:\C\else\if;\C\else\if-\C\else\if)\C\else\if/\C\else\if]\C\else\if'\C%
\else\space\fi\fi\fi\fi\fi\fi\fi\fi\fi\fi}%
\else\if\empty\C\else\if\space\C\else\space\fi\fi\fi}\fi}
\def\SP{\let\\=\empty\futurelet\C\puncspace }

\def\rf{$R_F$\SP }
\def\etal{et\SP al.\SP }

\def\deg{$^\circ$\ }

\def\kms{kms$^{-1}$ }
\def\h1{$h^{-1}$}

\begin{document}

\title{Two Galaxy Clusters:  A3565 and A3560\footnote {Based 
on observations at
Complejo Astronomico El Leoncito (CASLEO), operated under agreement between the
Consejo Nacional de Investigaciones Cient\'\i ficas de la Rep\'ublica
Argentina and the National Universities of La Plata, C\'ordoba and San
Juan; European Southern Observatory (ESO), partially under the ESO-ON
agreement and  Observat\'orio do Pico dos Dias, operated by the
Laborat\'orio Nacional de Astrof\'\i sica (LNA).} }

\author{
C.N.A. Willmer\altaffilmark{2,3,4,5}, 
M.A.G. Maia\altaffilmark{2,3,4},
S.O. Mendes\altaffilmark{3,6,7},
M.V. Alonso\altaffilmark{8}, 
L.A. Rios\altaffilmark{3,6},
O.L. Chaves \altaffilmark{3,4},
and D. F. de Mello\altaffilmark{4,9}}

\affil{Observat\'orio Nacional, Rua General Jos\'e Cristino
77, Rio de Janeiro, RJ 20921-030, Brazil} 

\altaffiltext{2}{Visiting Astronomer, CASLEO, Argentina}

\altaffiltext{3} {Visiting Astronomer, LNA, Brazil}

\altaffiltext{4} {Visiting Astronomer, ESO, Chile}

\altaffiltext{5} {Present address: UCO/Lick Observatory, University of
California, 1156 High St., Santa Cruz, CA 95064}

\altaffiltext{6} {Observat\'orio do Valongo, UFRJ, Ladeira do Pedro
Ant\^onio 43, Rio de Janeiro, Brazil}

\altaffiltext{7} {Present address: Department of Astronomy and Physics,
Saint Mary's University, Halifax, NS, Canada B3H 3C3}

\altaffiltext{8} {Observatorio Astron\'omico de C\'ordoba, Laprida 854,
C\'ordoba, 5000, Argentina}

\altaffiltext{9}{Present address: Space Telescope Science Institute,
which is operated by the Association of Universities for Research in
Astronomy, Inc. under contract to the National Science Foundation.}

\begin{abstract}
We report 102 new redshifts and  magnitudes for a sample of galaxies to \rf
$\sim$ 15.5 $mag$ in a 2.17\deg $\times$ 2.17\deg region centered on the
galaxy IC~4296, the most luminous member of the A3565 cluster.
Up to the limiting magnitude we find 29 cluster members, and 
measure a velocity dispersion of $\sigma$ = 228 \kms. The estimated
total mass for this system is $\sim$ 3.0 $\times$ \h1 10$^{13}$ M$_\odot$
(where $h$ = H$_0$/100 km s$^{-1}$Mpc$^{-1}$), and its
dynamical properties are quite typical of poor clusters
presenting X-ray emission.
We also find that galaxies with absorption lines are
more concentrated towards the center of the cluster, while systems
with emission lines are mainly located in the outer parts.
The small velocity dispersion of the cluster, coupled to
the known presence of an interacting pair of galaxies, and the large
extent of the brightest cluster galaxy, could indicate that galaxy
formation through mergers may still be underway in this system.

The surveyed region also contains galaxies belonging to the
Shapley Concentration cluster A3560. Within 30 arc min of the cluster
center, we detect 32 galaxies, for which we measure a  velocity dispersion of
588 \kms and a mass of  $\sim$ 2 $\times$ \h1 10$^{14}$
M$_\odot$. However, because our sample is restricted to galaxies brighter
than M$^*$, these values should be considered only as rough estimates.

\end{abstract}
\keywords{Galaxies: clusters: individual  -- galaxies: distances and redshifts
-- galaxies: photometry} 
\clearpage

\section{Introduction}

A region of the sky that presents a significant
number of groups and clusters is the so called Centaurus
Concentration (Lynden-Bell et al. 1989) which is a
collection of about 20 groups and poor clusters
which could be associated with the Great Attractor
(Lynden-Bell et al. 1988). To date, only a few of these
groups have been studied in detail (e.g., A3574 by Richter 1984; 1987;
Willmer et al. 1991; and S753, Willmer et al. 1991; J\o rgensen et
al. 1995), being overshadowed by the much richer Shapley Concentration
which lies in the background (e.g., Drinkwater et al. 1999; Bardelli
et al. 1998; Ettori et al. 1997 and references therein).

Another reason for the paucity of studies of
individual groups of galaxies in the region (as well as elsewhere) 
are the biases to which this
kind of system is susceptible. In general, when analysing 
groups identified from the wide-angle redshift surveys 
(e.g., Huchra \& Geller 1982; Maia et al. 1989; Nolthenius 1993; 
Garcia 1993; Ramella et al. 1999) one has to deal with
two potential problems. The first is the variable limiting absolute
magnitude of the group-finding algorithms (which is a function of
redshift), so that typically only the brightest components are
sampled. The second is the possibility that some of these groups are
only chance projections along the line of sight rather
than true physical associations, as only a few galaxies at most are
assigned to the groups. The most effective means of overcoming these
limitations is to probe the luminosity function of the groups
to much fainter limits (e.g., Zabludoff \& Mulchaey 1998a; Mulchaey \&
Zabludoff 1998a;  Koranyi et al. 1998; Mahdavi et al. 1999), which 
allows not only establishing which galaxies are group members, but
also ensures a more robust estimation of the group's physical
parameters (e.g., Zabludoff \& Mulchaey 1998a).

In this work  we present a redshift survey of a $\sim$ 4 square
degrees region centered on IC~4296, which is the brightest galaxy of
the cluster A3565, one of the groups belonging to
the Centaurus Concentration.
This cluster, located at R.A. 13$^h$ 33.8$^m$
and Dec. --33\deg 43$'$ (1950.0) is classified as a Richness 1 cluster by
Abell et al. (1989), and as a  BM type I cluster
by Brown \& Burns (1991). The group associated with IC~4296 was 
first identified by Sandage (1972), and has been included in the
group catalogues of Huchra \& Geller (1982) (containing 3 galaxies) and
Garcia (1993) (where it has 13). 
Most of the works dealing with this group have concentrated on
IC~4296, which is the optical counterpart of the low-luminosity
radio-source PKS~1333-33 (Mills et al., 1960; Killeen \& Bicknell 1988). This
galaxy also presents X-ray emission which has
been studied by  Forman et al. (1985) using {\it{Einstein}} data,
and more recently by Buote \& Fabian (1998) using {\it{ASCA}} data.
The diffuse X-ray emission of the cluster detected by {\it{ROSAT}} 
has been studied by  Mulchaey et al. (1996).
The properties of several of its brighter members were investigated in
a series of papers by Kemp \& Meaburn (1991, 1993) and Kemp (1994).
This cluster has also been used by Lauer \& Postman (1994) and
Lauer et al. (1998) in the study of peculiar velocities in the
local Universe, and to estimate the far field Hubble constant. In the
latter work, Lauer et al. (1998) have estimated the distance 
of IC~4296, the brightest cluster galaxy, as 3220 kms$^{-1}$.

The region surveyed in this work also includes A3560 (13$^h$29.0$^m$,
--32\deg 58$'$, 1950.0, Abell et al. 1989), who classified it as an
irregular  system of richness class 3, and estimated its redshift as
$cz \sim$ 3270 kms$^{-1}$, the central position corresponding to that of
the galaxy NGC~5193. In the analysis of the Shapley Concentration, 
Vettolani et al. (1990) noted that NGC~5193 and NGC~5193A are actually
galaxies in the foreground  of the cluster CE1329-327 identified by
Melnick \& Moles (1987), the center of which is 14$'$ off (mostly in
Declination ) from the Abell et al. (1989) position. From the measurement of
ten spectra Melnick \& Moles (1987) found that  $cz$ $\sim$ 14850
kms$^{-1}$, which places A3560  
in the Shapley Concentration. That A3560 is a background cluster was
also noted by Mould et al. (1991), though in their work they retain
the name of A3560 to identify a foreground group (which contains some
galaxies of A3565). Lauer et al. (1998) also refer to the distribution
of galaxies at $\sim$ 3800 \kms centered on NGC~5193 as A3560. In
this work we will assume that both NGC~5193 and NGC~5193A are
part of A3565, and
follow Vettolani et al. (1990) by calling A3560 the cluster
in the Shapley Concentration. Other works dealing with
this system have been Gregorini et al. (1994), who detected radio
emission and showed that its peak originates from a position close to a
dumbbell galaxy in the cluster; and Ebeling et al. (1996), Pierre et
al. (1994) and  Ettori et al. (1997), who analyse the X-ray emission
detected by {\it{ROSAT}}.

This paper is organized as follows: the acquisition and observations
of the galaxy sample are described in Section 2, 
followed by the analyses in Section 3.
A summary of the main results follows in Section 4.

\section{The Sample}

\subsection{ Acquisition, Astrometry and Photometry}
The sample of galaxies was defined from scans of an on-film copy of
the ESO/Uppsala survey field 383 in the $R$ band (IIIaF emulsion +
GG630 filter, Lauberts \& Valentijn 1989, hereafter ESO--LV) using the
Observat\'orio 
Nacional PDS 1010A microdensitometer. A square slit of 20$\mu m$ size was
used, corresponding to 1.35$''$ $\times$ 1.35$''$ projected on the sky,
about 0.25\h1 $\times$ 0.25\h1 kpc at the cluster distance, where
$h$ = $H_0$/100 \kms Mpc$^{-1}$. We scanned an area of 2.17\deg $\times$
2.17\deg (1.4\h1 Mpc $\times$ 1.4 \h1 Mpc) centered close to 
the position of IC~4296. The scanned area was divided into a mosaic of 9 scans
each of 2000 $\times$ 2000 pixels, with 100 pixels overlap between
contiguous scans. This was done to minimize possible variations of the
instrument focus due to the fact that we were not scanning a plate, while
the overlap ensured that few objects on the border between scans, if any,
would be lost, and would also allow a verification of the photometric
stability.
The scans were transformed into intensities in a procedure 
entirely analogous to that of Marston (1988) where the
characteristic curve is derived comparing the surface
brightness profiles of galaxies measured from the scans and
CCD photometry.
After the transformation into intensities, the scans were
processed with FOCAS (Jarvis \& Tyson 1981; Valdes 1982), and a possible
candidate  object was considered whenever it contained more than 9
contiguous pixels above the threshold, which we took as being 3
$\sigma$ above the sky level. Objects in common to 
one or more scans were later used to shift the scans
into the same instrumental magnitude system (e.g., Maddox et
al. 1990).

The astrometric solution for each of the 9 individual scans was obtained
independently. The first step was to identify in the object catalogue
stars contained in the HST Guide Star Catalogue (Lasker
et al. 1990, hereafter HSTGC). The positions of these stars
were redetermined following the procedure of Assafin et al. (1997),
using the PPM (R\"oser \& Bastian 1991) and ACRS (Corbin \& Urban 1988)
catalogues. 
Basically, we selected  PPM and ACRS stars in a 2\deg wide region
centered on the individual scan, which were used to refine the HSTGC
star positions, therefore defining the secondary
reference frame with a rms error of $\sim$ 0.5$''$. The secondary frame 
was then used to transform the scan $x$ and $y$ positions into right
ascension and declination. We estimate that positional uncertainties
should be smaller than 1$''$. 

The calibration used CCD photometry for 15 galaxies obtained during
two observing runs at the LNA, the details of which are described  
in Table 1. The magnitudes were measured
at three different isophotal levels, as described by Alonso et
al. (1993).
The calibrating galaxies are presented in Table 2
which shows in column (1) the galaxy identification
used in this work, followed in column (2) by the previous
identification when known; the $R$ magnitudes and errors for each of
the surface brightness levels 24, 25 and 26 $mag~arcsec^{-2}$ are
presented in columns (3), (4) and (5), and column (6) identifies the
observing run described in Table 1.
The magnitudes used in our catalogue, which we will denominate \rf,
were obtained through a linear fit between the $R_{25}$ and the
instrumental photographic magnitudes.
An estimate of the uncertainty of our \rf magnitudes was obtained
by comparing our measurements with $R_{25}$ of the ESO-LV catalogue for
25 galaxies in common. The comparison presented in Figure
1  shows that there is a mean difference 
\rf \SP--~ $R_{25}(ESO)$ = --0.05 $\pm$ 0.24 $mag$.
For the 18 galaxies we have in common with Drinkwater et al. (1999) we find
\rf \SP--~$R_{DPP}$ = 0.11 $\pm$ 0.25 $mag$. Both comparisons suggest
that our magnitudes have an average probable uncertainty of $\sim$
0.18~$mag$ if the errors are similar for each of the data sets. 

The final catalog used to define the spectroscopic observations was
limited at \rf = 15.5 $mag$ before making reddening corrections. In
the analyses in Section 3, galaxy magnitudes were corrected
for extinction using the  DIRBE--IRAS maps of Schlegel \etal (1998)
interpolated for the position of each galaxy, where we have assumed that 
$A_{R_{\scriptscriptstyle{F}}}$ = $A_R$ . In general the average
reddening in the $R$ band for this region is $\sim$ 0.15 $mag$.

\subsection{Spectroscopy}

The spectroscopic data used in this work were obtained at different
telescopes, which are summarized in Table 3, where we identify the
site (column 1), the telescope size (column 2), while the detector is
identified in column (3) and its size in column (4); column (5)
shows the grating rule, followed by the  dispersion (column 6),
resolution (column 7) and wavelength coverage (column 8).
The data were reduced in exactly the same manner as the SSRS2 data (da
Costa et al. 1998), following the usual procedures of removing bias
images, correcting by flatfield and making illumination corrections
(e.g., Massey 1992), all  within IRAF\footnote {IRAF is distributed by
the National Optical Astronomy Observatories (NOAO) which is operated
by the Association of  Universities for Research in Astronomy, Inc. under 
contract to the U. S. National Science Foundation}.
Radial velocities were measured using the RVSAO package
(Kurtz et al. 1992; Kurtz \& Mink 1998). In addition to the standard
templates supplied with RVSAO, we used composite spectra of stars of
various spectral types, a 
composite spectrum of galaxies measured for the SSRS (da Costa et al.
1989), and high signal-to-noise spectra of NGC 7507 and M31, the
latter kindly supplied by V. de Lapparent and C. Bellanger. As in the
case of the SSRS2 data (da Costa et al. 1998), which shared the same
instrumental setups of the majority of observing runs reported herein,
and was processed with the same software,
the estimated velocity error for most spectra should be of the order
of $\sim$ 40 \kms. An independent assessement of these errors can be
made by comparing redshifts for 18 galaxies in common with
Drinkwater et al. (1999), where we find a mean
difference $v$ ~--~ $v_{DDP}$ = 40 $\pm$ 87 \kms. This result implies in $\sim$
61 \kms uncertainty if both data sets have comparable errors, while if
we consider the average uncertainty quoted by Drinkwater et al. (1999)
for their measurements (67 \kms) this implies in an uncertainty of 55 \kms
for our radial velocities. There are 6 galaxies in common with 
Quintana et al. (1995) for which we also measured radial
velocities. We find a mean difference of  $v$~--~$v_{QRM}$ = 125 $\pm$
83  \kms. The origin of this fairly large offset is unclear.

The final sample of galaxies is presented in Table 4
\footnote{An electronic version of this table is available at
$http://www.dan.on.br/other\_surveys/a3565.html$}
where we give in
column (1) our object identification; in column (2) the
previous identification obtained by matching galaxy positions with
objects in the NED database, and for a few cases with the HSTGC, the
original reference being identified in the footnotes; 
the right ascension and declination for epoch B1950.0 in columns (3) and (4)
respectively and the \rf magnitudes in column (5). A rough
morphological classification is shown in column (6), in
a system consistent with that of da Costa et al. (1998), where
T= --5 corresponds to ellipticals; --2 to S0s; 0 to S0/a; 5 to spirals
in general; 16 to peculiar morphologies; 22 to dwarfs and 33 to
galaxies with superposed stars. We should note that for fainter
galaxies these classifications are uncertain because of the small
apparent sizes.
This is followed in column (7)  by
the heliocentric radial velocity; the estimated internal error (column 8); the
number of emission lines used in the redshift determination (column 9)
and the source code for the radial velocity (column 10), where s1,
s2, etc. refer to the run number in Table 3, or the references shown
in the table notes.
In the table we have included all galaxies identified in this
region to about \rf $\sim$ 16.4 $mag$.
A total of 166 new observations are reported in this work, including
the outer region discussed in Appendix A.

\section{Analysis}

\subsection {Distribution of Galaxies}

The sample of galaxies brighter than \rf = 15.5 $mag$ considered in this
work is shown in Figure 2, where the symbols represent galaxies 
in different redshift intervals, as noted in the caption, while the
symbol size is a function of the object magnitude.
The most prominent features in this plot are A3560 at approximately 
13$^h$ 30$^m$ and --32\deg 50$'$, at the upper right, and
A3565 at 13$^h$ 33$^m$ --33\deg 40$'$, close to the center of the surveyed region.
To the adopted magnitude limit, there are 111 galaxies, of which
110 (99\%) have measured radial velocities.
The distribution in radial velocities bins of 250 \kms
is shown in Figure 3, where all velocities were corrected for
galactic rotation through  $v_o = v_\odot + 300 \sin{l} \cos{b}$. The
two main peaks correspond to A3565 at $\sim$ 3600 
\kms and galaxies associated with A3560 and the Shapley Concentration 
at $\sim$ 15000 kms$^{-1}$. The smaller peak at $\sim$ 7000 \kms is due
to a group of 3 galaxies close to the position of IC~4296, plus a few
other galaxies distributed over the whole field. This loose
distribution of galaxies at $\sim$ 7000 \kms has also been detected by
Drinkwater et al. (1999). In contrast, the structure at $\sim$ 11000
\kms found by those authors in the foreground of the Shapley
Concentration is not very prominent in our sample; an inspection
of their redshift maps suggests that most of the galaxies in that
structure are located to the south of our survey.

\subsection { A3565}

The peak corresponding to A3565 in Figure 3 contains 30 galaxies 
which are well separated in velocity space from the foreground galaxies 
at $\sim$ 2000 \kms and the background at $\sim$ 7000 kms$^{-1}$.  
Because the surveyed area is rather small ($r < 0.7$ \h1 Mpc), it is
possible that some outlying galaxies are being lost (e.g., Mahdavi et
al. 1999). This in fact is suggested when comparing the present
catalogue with that derived 
by Garcia (1993) using the LEDA database (Paturel et al. 1991), which
contains three galaxies beyond the survey boundaries. However, in
order to keep our sample homogeneous and well defined, we have not
considered in the analysis these possible member galaxies.
In this work we will consider as group members only those galaxies
that are within the surveyed area and with radial velocities
within 3 $\sigma$ of the cluster mean (Yahil \& Vidal 1977), where
$\sigma$ is the cluster velocity dispersion uncorrected for velocity
errors. This eliminates WMMA 219 ($v_\odot$=4494 \kms) from the sample.
The spatial distribution of
likely member galaxies of A3565 is shown in Figure 4, where again we
have coded the symbol size as a function of the apparent \rf magnitude
of the galaxy, while crosses represent galaxies with emission lines
in the spectra, and open squares galaxies only presenting absorption
line spectra.

In Table 5 we present the dynamical properties of A3565, most
of which were calculated using the ``classical
estimators'' (mean radial velocity, velocity dispersion and mass),
while the cluster centroid was estimated using an unweighted average.
The velocity dispersion and errors were derived following 
Danese et al. (1980), where the radial velocity uncertainty is
accounted for, while for the other parameters we used the
``jacknife'' technique (e.g., Bothun et al. 1983). In the table
we also show the values of the central location and scale
estimates calculated using ROSTAT (Beers et al. 1990), the
errors for the latter being estimated by means of 1000 bootstrap
simulations. As may be seen in the table, the values of both
classical and robust estimators agree very well.
All these parameters were calculated without weighting
by luminosity. The mean harmonic
radius (R$_h$), the mean pairwise separation (R$_p$) and
the virial mass (M$_{VT}$) were calculated using the expressions of
Ramella, Geller \& Huchra (1989). The uncertainty for the
latter was estimated using standard error propagation, which takes
into account the uncertainty in $\sigma$ and the jacknife error for
R$_h$ . We also calculate the projected mass
(M$_p$) which is an estimator proposed by Heisler, Tremaine \& Bahcall (1985) 
that is less sensitive to the presence of interlopers. Both
M$_{VT}$ and M$_p$ were shown by West, Oemler \& Dekel (1988) to give
reliable measures of the cluster mass.

The velocity dispersion of A3565 (228 \kms) is marginally
higher than the typical value for groups of galaxies 
$\sim$ 194 \kms (Ramella et al. 1999), yet
in the lower range  of velocity dispersions measured by Zabludoff \&
Mulchaey (1998a) in their sample of 12 poor clusters with X-ray emission.
The mass derived from the Virial Theorem in addition to the projected mass
estimator by Heisler et al. (1985) are in good agreement, giving a
mass of the order of 3 $\times$ 10$^{13}$ M$_\odot$. The uncertainties
for M$_{VT}$ are of the order of 30\%, a value which is consistent
with that measured for simulated clusters by West et al. (1988). The
estimated uncertainty measured for M$_p$ is probably too small, as the
latter authors found that it presents comparable errors to those
of M$_{VT}$.

The dynamics of A3565 is significantly influenced by the
presence of IC~4296, which, although not classified as a cD,
does present an extended profile quantified by a value of n=11.8 for a
Sersic law (Graham et al. 1996). Such flat profiles were noted by
those authors as being typical of Brightest Cluster Galaxies.
The large extent of this galaxy was also noted by Saglia et al. (1993)
who presented the velocity dispersion profile out to a distance of 0.8
R$_e$ which corresponds to 45$''$ or 7.7 \h1 kpc. The internal
velocity dispersion profile of this galaxy is greater than 200 \kms
and has a rather flat appearance, indicating the presence of a massive
dark halo (Saglia et al. 1993). 

The velocity histogram of group members (corrected to the
Local Group Centroid) is shown in Figure 5 where the dashed line indicates the
radial velocity of IC~4296 ($v_o$ = 3593 \kms). This is almost identical
to the value of the bi-weight estimate for the central location
velocity of the cluster (3586 \kms). This result differs from the
conclusion of Kemp (1994), who found that a difference of  $\sim$ 150
\kms between IC~4296 and the group's mean radial velocity, though the
reason for this discrepancy is unclear.
The projected separation of IC~4296 to the group's centroid about 0.14 \h1 Mpc.
In order to check whether this could be due to the presence 
of interlopers at large angular separations, we also determined the
dynamical parameters in
the case where only galaxies within a radius of $\sim$ 0.5\deg 
(0.31 \h1 Mpc) from
IC~4296 were considered. This radius excludes galaxies such as
NGC~5193 and NGC~5193A which, as mentioned in the Introduction, have
been assigned to another system by Abell et al. (1989) and Lauer et
al. (1998). This smaller sample contains 21 galaxies, and
reduces the projected separation between the group center and IC~4296
to 0.07 \h1 Mpc, while the radial velocity difference is slightly
larger, increasing from 7 \kms to 39 \kms, still within the
measurement errors. The velocity dispersion for this smaller sample
also changes slightly (234 \kms). The mass estimators are $\sim$ 10\% smaller 
(M$_{VT}$~=~2.6~$\times~10^{13}$~M$_\odot$ ; 
M$_p$~=~3.0~$\times10^{13}$~M$_\odot$), but still within the errors
estimated from the larger sample, suggesting that the larger sample
probably contains no interlopers.

The spatial segregation between galaxies presenting emission lines 
(crosses), comprising 34\% of the sample (10 galaxies), from the
19 (66\%) galaxies that only present absorption line spectra (open squares),
can be seen in Figure 4. The latter tend to be found in the central
regions of the cluster, while emission line galaxies are predominant 
in the outer regions, a feature which is common in rich clusters, but
only recently shown to occur also in
poor clusters (Zabludoff \& Mulchaey 1998a; Mahdavi et al. 1999).
The observed segregation may be quantified through R$_h$ and
R$_p$ which for the absorption line galaxies are
R$_h$=0.28$\pm$0.02\h1~Mpc, and  R$_p$=0.47$\pm$0.04\h1~ Mpc, while for
the emission line systems these are R$_h$ = 0.67 $\pm$
0.03 \h1 Mpc and R$_p$ = 0.66 $\pm$ 0.10 \h1 Mpc, which are
a factor of $\sim$ 2 larger than for the absorption line systems. For
the entire sample the values are R$_h$ = 0.40 $\pm$ 0.01 
\h1 Mpc and R$_p$ = 0.57 $\pm$ 0.03 \h1 Mpc.
By running a two-sample Kolmogorov-Smirnov test on the projeted
distances of galaxies from the cluster center we find that the probability of
both samples being drawn from the same parent population is at the 4\%
level; a similar result ( 3.6\%) is obtained from Monte-Carlo simulations
where the galaxy distances from the cluster center
were bootstrapped. On the other hand the cluster velocity
dispersion measured using either type of galaxy is to all effects
identical. For the 
19 absorption line systems we find 231 (+51, --32) \kms using the
classical estimator, and 242 $\pm$ 42 \kms using the
bi-weight ROSTAT estimator, while for the 10 emission line systems
these are 232 (+82, --41) \kms and 235 $\pm$ 66, respectively.
The values for both galaxy populations are very similar to the
velocity dispersion of the entire sample 228 (+38, --26)\kms and
236 $\pm$ 69 \kms.

The distribution of absolute magnitudes is presented in Figure 6, 
where galaxies are counted in 0.5 magnitude bins. The faintest
absolute magnitude that can be reached with this sample is $\sim$ 3
magnitudes fainter than the value of M$_R^{*}$ (--20.29 $\pm$ 0.02
+ 5 log $h$, Lin et al. 1996). 
In the figure we may see a dip in the absolute magnitude distribution
at $M_R$ $\sim$ --19. This is at a similar value to that found by
Koranyi et al. (1998) in AWM 7 as well as in other systems, and which
could be a common feature to rich clusters, due to a depletion of
galaxies as a result of mergers (Koranyi et al. 1998).
However, because of the small number of galaxies in A3565, we cannot
consider this feature as being significant as it could be just a
fluctuation due to small number statistics.

In the study of groups and poor clusters presenting diffuse X-ray
emission detected by {\it{ROSAT}}, Mulchaey et al. (1996) present in
their Figure 4 the X-ray contours overlaid on a Digital Sky
Survey map covering a 1\deg $\times$ 1\deg region centered on IC~4296.
There one can see that the X-ray emission is not centered on
IC~4296, but presents
a peak between IC~4296 and IC~4299, having a rather irregular shape
aligned NE-SW.  By fitting a modified King model
\begin{equation}
S(r) = S_0 \lbrace 1 + (r/R_{core})^2 \rbrace^{-3 \beta + 0.5} + B_0
\end{equation}
to the counts, Mulchaey et al. (1996) measured $\beta$ = 0.46 and a
temperature $T$ = 1.07 keV. In their model, Mulchaey et al. (1996) 
considered the gas
profile out to a maximum extent of  21.5 arc min or 0.23 \h1 Mpc from
IC~4296, where the gas reaches 20\% of the background level.
The mass estimated from the X-ray model is of the order of 1.17
$\times$ 10$^{13}$ M$_\odot$, which is about a factor of 3 smaller
than the mass we derive from the galaxy distribution. 
By extrapolating the model of Mulchaey et al. (1996) 
out to the value of the harmonic radius measured
in this work we find that the mass estimate (1.57 $\times$ 10$^{13}$
$M_\odot$)  is still only about a half of the total mass
obtained from the virial and projected mass estimators. 

As noted above, IC~4296 has an internal velocity dispersion $\sim$ 200
\kms, which reaches up to $\sim$ 300 \kms in its central regions
(Wegner et al. 1999; Kemp 1994). This value is comparable to the
velocity dispersion of A3565, and it is interesting to note that of
the sample of 12 clusters studied by Zabludoff \& Mulchaey (1998a),
only 2 present a similar feature, although the significance of this
result is unclear. 
The high value of the velocity dispersion could imply that at least
some of the X-ray emiting gas is more influenced by the galaxy rather
than the cluster (e.g., David \& Blumenthal 1992).
The scenario where gas is bound by the central galaxy as well as by
the cluster potential is consistent with the conclusions of
Mulchaey \& Zabludoff (1998a), who obtained better fits to the 
observed X-ray distribution when a two-component model was considered:
one component would be due to the central galaxy and
the second component due to the diffuse gas. Such a scenario was
considered by Mulchaey et al. (1996), to explain the rather low value
of $\beta$ they measured for A3565.

\subsection { A3560}

In Figure 7 we show the distribution of galaxies with radial
velocities between 13000 \kms $\leq v_\odot \leq$ 16000 kms$^{-1}$, which
roughly corresponds to the redshift interval of the Shapley
Concentration. In addition to A3560 there is a dispersed 
distribution of galaxies which belong to the Shapley Supercluster,
but are not part of the cluster. 
Even though A3560 is close to the NW border of the surveyed region,
enough galaxies are contained in the survey that an
estimate of the dynamical parameters of the cluster is possible.
In this analysis we will consider the galaxies
within the range of 13$^h$ 28$^m  < \alpha <$ 13$^h$ 33$^m$ and 
-32\deg 35$' > \delta >$ -33\deg 20$''$ as likely members. Within
these limits and the radial velocity range above, there are 33
galaxies.
Their distribution seen in Figure 7 suggests
that A3560 could be composed of two concentrations of galaxies. This
is also suggested by the histogram of radial velocities in Figure 8,
where two velocity peaks are seen: one at $\sim$ 13600 \kms
and the other at $\sim$ 14800 kms$^{-1}$. 
In order to confirm whether both the spatial and redshift distribution
could be due to subclustering, we applied the statistical tests
described by Pinkney et al. (1996) to the sample. Only  
the  Lee 2-D and Lee 3-D statistics detect any significant level of
subclustering. The Dressler \& Shectman (1988) $\Delta$ statistic
in particular, which
is the most sensitive test (Pinkney et al. 1996), detects no
significant evidence of subclustering. Therefore, in the determination
of the physical parameters for A3560, shown in the last column of Table 5,
we will assume that there is no subclustering. In the analysis that
follows we also applied the  3 $\sigma$ clipping algorith (Yahil \&
Vidal 1977), which removes one galaxy from the sample (WMMA 076,
$v_\odot$ = 13361 \kms).

The velocity dispersion we measure for this cluster, (588 kms$^{-1}$) is
smaller than that measured by Melnick \& Moles (1987), 
838 kms$^{-1}$ which was based on 10 galaxies. Our mass estimate using
 the  Virial Theorem
($\sim$ 2 $\times$ 10$^{14}$ M$_\odot$) is close to the value
estimated by Ettori et al. (1997) from X-ray data ($\sim$ 2.3
$\times$ 10$^{14}$ M$_\odot$). The projected mass is a factor of 2
larger than the Virial estimator, and this could be due to
the combination of the small sample size with a somewhat complex
distribution of galaxies, which is seen not only in the optical
data, but also in the X-ray maps for this cluster of Pierre et
al. (1994). As may be seen in their Figure 6, there are four X-ray peaks
in the direction of A3560, and excepting for the most extended source,
they are not well correlated with the distribution of galaxies down to
the limiting magnitude used in this work.

We should note that the centroid we determine here for A3560
differs from the position of
Melnick \& Moles (1987), probably because of the larger range in
right ascension we are considering. The range used here corresponds to $\sim$
1.6 \h1 Mpc at the mean cluster distance. The centroid in Table 5 is
about 0.13\deg, or 0.32 \h1 Mpc away from the position of galaxies
WMMA 032 and WMMA 033 in Table 4, which form a dumbbell system (Gregorini et
al. 1994). The fact that this
type of system is typically found in the central regions of clusters
(Valentijn \& Casertano 1988), suggests that the centroid of A3560 may
be close to this position.  Further support for this interpretation
is the presence of the main component of the diffuse X-ray emission
detected by Pierre et al. (1994) which also peaks close to the
position of the dumbbell galaxy. In particular, if one considers the
sample of galaxies to \rf=16.0 $mag$, which has 80\% radial velocity
completeness, the distribution of galaxies with redshifts centered on
the dumbbell becomes much more pronounced. 
Furthermore, the fact that the mass-to-light ratio is large
($\sim$ 760 (M/L$_R$)$_\odot$) is also suggestive that the cluster
membership as considered in this work is uncertain.
All these results suggest that our conclusions regarding A3560 should only be
considered as preliminary, since a larger and fainter sample will be
required to describe more accurately the distribution of galaxies
in this region.

The radial velocities of the dumbbell components differ by 10 \kms,
while their projected separation is of the order of 9 \h1 kpc. Both
values are somewhat smaller that the typical values for this kind of
system quoted by Valentijn \& 
Casertano (1988). Both nuclei have absorption line spectra, neither
presenting any significant evidence of having undergone recent
episodes of star formation.  The mean radial velocity of these
systems is shown in Figure 8 as a dotted line, which is about 60 \kms
away from the mean radial velocity of the group in Table 5.

\section{Conclusions}

In this paper we have presented the redshift survey of a
2.17\deg $\times$ 2.17\deg region centered on the galaxy IC~4296,
which is the brightest galaxy in the A3565 cluster. 
The present work extends the coverage of galaxies of A3565 
to a limit of $M_R^*$+ 3, within a radius of $\sim$ 0.7 \h1 Mpc. Both
the radial extent and the number of members for this
cluster are comparable to what was found for similar systems
by Zabludoff \& Mulchaey (1998a) and Mulchaey \& Zabludoff (1998a) in 
their study of the optical and X-ray properties of 12 poor clusters
of galaxies. The velocity dispersion, as well as the mass we measure
for A3565 are within the range of values quoted in those works, and
A3565 has dynamical properties typical of X-ray emitting poor clusters. 

This cluster contains  galaxies undergoing interactions (NGC~5215A and
NGC~5215B; Kemp \& Meaburn 1991), and with common haloes (IC~4296 and
IC~4299; Kemp 1994). This evidence coupled to the small value of
the velocity dispersion of the system suggests that bright galaxies
could still be forming in A3565 through mergers.
Such a scenario has been suggested in recent works by Mulchaey \&
Zabludoff (1998b) and Zabludoff \& Mulchaey (1998b) who find that the
brightest galaxies in groups as well as the groups themselves could
still be forming through the accretion of smaller systems, and that
the merger rate is most efficient in groups with $\sigma \sim$ 200
\kms, which is the case of A3565. 

As in other poor clusters with diffuse X-ray emission,
(e.g., Zabludoff \& Mulchaey 1998a; Mahdavi et al. 1999),
A3565 also shows radial segregation between galaxies presenting
emission lines, mainly found in the outer parts of the cluster, while
absorption line systems are concentrated towards the cluster
center. This radial segregation together with the bending of the
radio jets of IC~4296 noted by Killeen \& Bicknell (1988), and
the presence of several galaxies with distorted morphologies (Kemp \&
Meaburn 1993), suggest that the Intergalactic Medium of A3565
could also play an important role in the dynamical evolution of this
system.

We also measured redshifts of galaxies belonging to the A3560 cluster, a
background system that belongs to the Shapley Concentration.
Although the velocity dispersion we measure for A3560 ( $\sim$ 580 \kms)
is typical of rich galaxy clusters, our results for this cluster
should only be taken as preliminary. This is suggested by
the relatively
bright limiting absolute magnitude of our work at the distance of
A3560 (M$_{lim} \sim$ M$_R^*$); the relatively complex distribution of
galaxies and X-ray emission; the discrepancies between the mass
estimators, and the large M/L ratio we measure.
Therefore, to characterize the membership and dynamics of this
cluster, a larger and fainter sample will be needed.

\appendix
\section{Additional Galaxies}
A more extensive though shallower survey in a larger (5\deg $\times$
5\deg) region in the general direction of A3565 shows no significant 
increase in the number of cluster members.  A further 51 galaxies
were observed, and are listed in Table 6 where
column (1) shows the HSTGC identification, and in column (2) other
identifications if known; columns (3) and (4) the right
ascension and declination for epoch B1950.0 followed  by the new
radial velocity and the internal estimated error, respectively in
columns (5) and (6).  In this table we also include three
serendipitous objects that were observed in the course of the survey,
which are noted in the table. The coordinates for these three objects
were estimated from the DSS.
Because most objects in Table 6 were defined from a blue selected sample
with $m_B~\leq$~15.5 $mag$, where $m_B$ is the blue magnitude at the 26
$mag~arcsec^{-2}$ (Alonso et al. 1993; da Costa et al. 1998), we did
not include them in the analysis of the group. However, the distribution of
galaxies may be seen in Figure 9, where we code galaxies in different redshift
intervals as in Figure 4. The main feature to note is that very few
galaxies in the redshift interval of A3565 are added in the outskirts
of the cluster.

\acknowledgments
We would like to thank Marcelo Assafin for his help on the astrometric
calibration, Jason Pinkney for providing the code to calculate
significance levels of subclustering, and Ricardo Schiavon for
the observations in the s11 run. We also thank Ann Zabludoff,
David Buote, George Blumenthal and Hern\'an Muriel for useful
conversations.
Financial support for this work has been given through FAPERJ (CNAW,
MAGM, OLC, DFM), CNPq grants 301364/86-9
(CNAW), 301366/86-1 (MAGM), 301456/95 (DFM); NSF AST 9529098 (CNAW);
and PICT97 No. 03-00000-01213 (FONCyT), CONICOR, SECyT, and CONICET
(MVA) and Fundaci\'on Antorchas--Vitae--Andes cooperation.
LAR and SOM were supported by CNPq undergratuate studentships offered
by the CNPq/ON through the PIBIC program. This research has made
use of the NASA/IPAC Extragalactic Database (NED) which is operated by
the Jet Propulsion Laboratory, CALTECH, under contract with the
National Aeronautics and Space Administration. We acknowledge the use
of NASA's $SkyView$ facility (http://skyview.gsfc.nasa.gov) located at
NASA Goddard Space Flight Center, and the use of the CCD and
data acquistion system supported under U.S. National Science
Foundation grant AST-90-15827 to R.M. Rich. We acknowledge use of the
Digitized Sky Survey, produced at the Space Telescope Science Institute
under U.S. Government grant NAG W-2166. The images are based on
photographic data obtained using UK Schmidt Telescope, operated by the
Royal Observatory Edinburgh, with funding from the UK Science and
Engineering Research Council (later the UK Particle Physics and
Astronomy Research Council), until 1988 June, and thereafter by the
Anglo-Australian Observatory.

\begin{deluxetable}{cccccccc}
\tablewidth{38pc}
\tablecaption{Observations: Photometry}
\tablehead{
\colhead{code} & \colhead{date} & \colhead{telescope} & \colhead{CCD} & \colhead{Format} &
\colhead{Scale} &
\colhead{Gain} &
\colhead{RON} \nl
{} & {}     & {}     & {}    &  {} &
[$''/pix$] & [e$^-$/ADU] &  [e$^-$] \nl
(1) & (2) & (3) & (4) & (5) & (6) & (7) & (8)
}
\startdata
$p$1 & 1996 March 21 & 0.60 m & LNA\#301 & 578 $\times$ 385  & 1.13 & 2.63 & 5.4 \nl
$p$2 & 1997 May 2    & 1.60 m & LNA\#48  & 770 $\times$ 1152 & 0.57 & 5.59 & 10.67 \nl
\enddata
\end{deluxetable}

\begin{deluxetable}{llrrrl}
\small
\tablewidth{0pc}
\tablecaption{Galaxies in Scans with CCD Photometry}
\tablehead{
\colhead{WMMA Id.}       &
\colhead{Identification} &
\colhead{$R_{24}$}       &
\colhead{$R_{25}$}       &
\colhead{$R_{26}$}       & 
\colhead{Run}            \nl
~~(1) &~~~~~~~(2) & (3)~~~~~~ & (4)~~~~~~ & (5)~~~~~~ & (6)
}
\startdata
008 & NGC~5193A,~383~G~14 & 12.98 $\pm$ 0.22 & 13.05 $\pm$ 0.18 & 13.11 $\pm$ 0.13 & p1 \nl
011 & NGC~5193,~383~G~15 & 11.27 $\pm$ 0.17 & 11.17 $\pm$ 0.31 & 11.14 $\pm$ 0.40 & p1 \nl
125 &  {}           & 16.15 $\pm$ 0.05 & 16.06 $\pm$ 0.06 & 15.98 $\pm$ 0.08 & p1 \nl
126 & 383G37        & 13.24 $\pm$ 0.05 & 13.16 $\pm$ 0.06 & 13.11 $\pm$ 0.07 & p1 \nl
135 & B133341.2-334911 & 14.01 $\pm$ 0.05 & 13.97 $\pm$ 0.05 & 13.93 $\pm$ 0.06 & p1 \nl
137 & IC~4296,~383~G~39 & 10.24 $\pm$ 0.05 & 10.08 $\pm$ 0.06 &  9.93 $\pm$ 0.06 & p1 \nl
141 & IC~4299,~383~G~42 & 12.18 $\pm$ 0.05 & 12.16 $\pm$ 0.05 & 12.13 $\pm$ 0.05 & p1 \nl
159 & 383~G~45        & 12.48 $\pm$ 0.05 & 12.43 $\pm$ 0.05 & 12.38 $\pm$ 0.05 & p1 \nl
171 & 383~G~49        & 13.11 $\pm$ 0.06 & 12.84 $\pm$ 0.06 & 12.57 $\pm$ 0.06 & p2 \nl
192 & 383~G~55        & 14.40 $\pm$ 0.06 & 14.22 $\pm$ 0.06 & 14.03 $\pm$ 0.06 & p2 \nl
204 &    {}         & 15.89 $\pm$ 0.06 & 15.74 $\pm$ 0.06 & 15.58 $\pm$ 0.06 & p2 \nl
206 & B133737.8-334408 & 16.09 $\pm$ 0.06 & 15.94 $\pm$ 0.06 & 15.83 $\pm$ 0.06 & p2 \nl
208 &   {}          & 15.32 $\pm$ 0.06 & 15.02 $\pm$ 0.06 & 14.71 $\pm$ 0.06 & p2 \nl
209 &   {}          & 14.53 $\pm$ 0.06 & 14.32 $\pm$ 0.06 & 14.11 $\pm$ 0.06 & p2 \nl
231 &   {}          & 14.68 $\pm$ 0.05 & 14.49 $\pm$ 0.05 & 14.31 $\pm$ 0.05 & p2 \nl
\enddata
\end{deluxetable}

\begin{deluxetable}{rlllcrrrr}
\tiny
\tablewidth{38pc}
\tablecaption{Observations: Spectroscopy}
\tablehead{
\colhead{code} &
\colhead{date} &
\colhead{telescope} &
\colhead{CCD} &
\colhead{Format} &
\colhead{Grating} &
\colhead{Dispersion} &
\colhead{Resolution} &
\colhead{$\lambda$ coverage} \nl
{} & {}     & {}     & {}    & {} & [l/mm]~ & [$\AA/pix$]~~ &
[$\AA$]~~ & [$\AA$]~~~~ \nl
(1)~~ & ~~~(2) & ~~~~~(3) & ~~~~(4) & (5) & (6)~~ & (7)~~ & (8)~~ & (9)~~~~
}
\startdata
s1  & 1994 May & CASLEO 2.15m & EEV P8603S &  385 $\times$  578 & 600 & 2.4 & 7.3 & 4650--6050 \nl
s2  & 1994 Jul & CASLEO 2.15m & EEV P8603S &  385 $\times$  578 & 300 & 4.7 & 14.0 & 4100--6800 \nl
s3  & 1995 Mar & OPD 1.60m    & EEV P8603A & 1156 $\times$  770 & 900 & 1.2 & 3.1 & 4770--6135 \nl
s4  & 1995 Apr & CASLEO 2.15m & Tek        & 1024 $\times$ 1024 & 600 & 1.6 & 5.0 & 5100--6800 \nl
s5  & 1997 Jan & ESO 1.52m    & Loral      & 2048 $\times$ 2048 & 600 & 1.7 & 4.3 & 3600--7500 \nl
s6  & 1997 Apr/May & ESO 1.52m    & Loral  & 2048 $\times$ 2048 & 600 & 1.7 & 4.3 & 3600--7500 \nl
s7  & 1997 Jun & ESO 1.52m    & Loral      & 2048 $\times$ 2048 & 600 & 1.7 & 4.3 & 3600--7500 \nl
s8 & 1998 Feb/Mar & ESO 1.52m& Loral      & 2048 $\times$ 2048 & 600 & 1.7 & 4.3 & 3600--7500 \nl
s9 & 1998 Apr & ESO 1.52m    & Loral      & 2048 $\times$ 2048 & 600 & 1.7 & 4.3 & 3600--7500 \nl
s10 & 1998 Jun & ESO 1.52m    & Loral      & 2048 $\times$ 2048 & 600 & 1.7 & 4.3 & 3600--7500 \nl
s11& 1999 Feb & ESO 1.52m    & Loral      & 2048 $\times$ 2048 & 600 & 1.7 & 4.3 & 3550--7450 \nl
\enddata
\end{deluxetable}


\begin{deluxetable}{llrrrrrrrrr}
\tiny
\tablewidth{0pc}
\tablecaption{Galaxies in the Scanned Region}
\tablehead{
\colhead{WMMA Id.}                & \colhead{Identification}      &
\colhead{R.A.}                   & \colhead{Dec.}           &
\colhead{\rf}                    &  \colhead{T}              & 
\colhead{$v_\odot$}              &\colhead{$\pm$}           &
\colhead{N$_e$}                  & \colhead{Ref}         \nl
{}  & {} &  B1950.0    &   B1950.0     & {} & {} & \kms  & \kms& {} & {} & {} \nl 
~~(1) & ~~~~~~(2) & (3)~~~ & (4)~~~ & (5)~~ & (6) & (7)~~ & (8)~ & (9) & (10)  
}
\startdata
001 & NGC~5188,~383~G~09 & 13:28:39.6 & -34:32:19 & 11.03 &  5 &  2425 &6& 0 & T98\nl
002 &                 & 13:28:40.4 & -34:15:48 & 15.84 &  5 &   --  &  -- & -- & -- \nl
003 &                 & 13:28:47.6 & -34:06:48 & 15.49 &  5 & 13798 &  40 & 3 & s11 \nl
004 &                 & 13:28:49.6 & -32:59:09 & 14.43 &  5 & 16554 &  40 & 0 & s2 \nl
005 & 383~G~12          & 13:28:53.1 & -33:07:23 & 13.89 &  5 &  7759 &  21 & 1 & s9 \nl
006 &                 & 13:28:53.1 & -32:56:21 & 15.84 & -5 & 15449 &  35 & 0 & s8 \nl
007 &                 & 13:28:56.4 & -32:38:12 & 14.10 & -2 & 14780 &  46 & 0 & s9 \nl
008 & NGC~5193A,~383~G~14 & 13:28:58.9 & -32:58:52 & 13.30 & -2 &  3519 &  21 & 0 & s6 \nl
009 & GDP~1\tablenotemark{a} & 13:29:01.2 & -32:56:18 & 15.35 & -5 & 14852 &  40 & 0 & s8 \nl
010 &                 & 13:29:02.0 & -33:23:48 & 16.13 & -2 &   --  &  -- & -- & -- \nl
011 & NGC~5193,~383~G~15 & 13:29:03.1 & -32:58:39 & 11.69 & -5 &  3745 &  20 & 0 & s8 \nl
012 &                 & 13:29:08.9 & -33:47:07 & 15.45 &  5 & 14246 &  49 & 4 & s10 \nl
013 &                 & 13:29:09.0 & -33:47:45 & 16.28 &  5 &   --  &  -- & -- & -- \nl
014 &                 & 13:29:09.6 & -32:50:12 & 16.09 &  5 &   --  &  -- & -- & -- \nl
015 &                 & 13:29:10.9 & -32:38:20 & 16.28 &  5 &   --  &  -- & -- & -- \nl
016 &                 & 13:29:12.8 & -32:48:34 & 15.60 & -2 &   --  &  -- & -- & -- \nl
017 &                 & 13:29:13.1 & -32:37:29 & 15.27 &  5 & 14033 &  28 & 0 & s10 \nl
018 & B132914.5-333157\tablenotemark{b} & 13:29:14.4 & -33:31:58 & 15.39 &  5 &  8755 &  25 & 4 & s5 \nl
019 & B132916.4-330338\tablenotemark{b} & 13:29:16.5 & -33:03:41 & 15.06 & -5 & 13820 &  33 & 4 & s5 \nl
020 &                 & 13:29:17.4 & -32:45:13 & 14.65 & -2 & 14807 &  38 & 0 & s2 \nl
021 &                 & 13:29:18.8 & -33:07:46 & 15.51 & -2 &   --  &  -- & -- & -- \nl
022 &                 & 13:29:19.0 & -33:03:32 & 15.46 &  5 &  7539 &  58 & 8 & s5 \nl
023 & AM~1329-324\tablenotemark{c}     & 13:29:19.6 & -32:40:35 & 15.09 & -5 & 14214 & 106 & 0 & QRM \nl
024 &                 & 13:29:19.7 & -34:06:19 & 16.35 &  5 &   --  &  -- & -- & -- \nl
025 &                 & 13:29:19.9 & -33:20:32 & 16.17 & -2 &   --  &  -- & -- & -- \nl
026 & AM~1329-324\tablenotemark{c}     & 13:29:19.9 & -32:40:22 & 15.09 & -2 & 14801 &  49 & 0 & s10 \nl
027 &                 & 13:29:22.4 & -32:52:14 & 15.54 & -2 & 14541 &  30 & 0 & s11 \nl
028 & B132924.5-330715\tablenotemark{b} & 13:29:24.6 & -33:07:15 & 14.76 &  5 & 15292 &  43 & 0 & s2 \nl
243 &                 & 13:29:27.5 & -32:53:17 & 16.64 & -2 & 13502 &  51 & 0 & s11 \nl
029 &                 & 13:29:28.6 & -32:39:48 & 15.20 & -2 & 15703 &  29 & 0 & s5 \nl
030 & HSTGC~07269-01549\tablenotemark{d}     & 13:29:29.7 & -32:53:45 & 15.19 & -2 & 14868 &  26 & 0 & s10 \nl
031 & QRM~1325-32~41\tablenotemark{e},~B132933.4-325855\tablenotemark{b} & 13:29:33.4 & -32:58:55 & 14.89 & -2 & 13880 &  39 & 0 & s5 \nl
032 & QRM~1325-32~37\tablenotemark{e}          & 13:29:34.5 & -32:52:53 & 15.29 & 16 & 14734 &  32 & 0 & s11 \nl
033 & QRM~1325-32~38\tablenotemark{e}          & 13:29:35.3 & -32:52:44 & 14.44 & 16 & 14724 &  25 & 0 & s11 \nl
034 &                 & 13:29:37.0 & -32:50:50 & 15.80 &  5 &   --  &  -- & -- & -- \nl
035 &                 & 13:29:39.1 & -32:59:35 & 14.93 &  5 & 14592 &  51 & 0 & s7 \nl
036 & QRM~1325-32~42\tablenotemark{e}          & 13:29:40.1 & -32:54:20 & 15.72 & -2 & 14595 &  46 & 0 & QRM \nl
037 &                 & 13:29:41.0 & -32:54:55 & 16.15 & -2 &   --  &  -- & -- & -- \nl
038 & QRM~1325-32~40\tablenotemark{e}         & 13:29:43.1 & -32:53:53 & 14.72 & -5 & 12516 &  26 & 0 & QRM \nl
039 &                 & 13:29:44.0 & -33:02:47 & 16.05 &  5 &   --  &  -- & -- & -- \nl
040 & AM~1329-325\tablenotemark{c}     & 13:29:46.2 & -32:51:30 & 15.09 & -2 & 15410 &  29 & 0 & s5 \nl
041 & AM~1329-325\tablenotemark{c}     & 13:29:46.3 & -32:51:30 & 14.94 & -2 & 15031 &  48 & 0 & s5 \nl
042 &                 & 13:29:50.9 & -34:38:31 & 16.29 &  5 &   --  &  -- & -- & -- \nl
043 &                 & 13:29:54.5 & -32:58:58 & 15.33 &  5 & 14222 &  32 & 0 & s10 \nl
044 &                 & 13:29:55.6 & -33:08:46 & 15.84 &  5 &   --  &  -- & -- & -- \nl
045 &                 & 13:29:59.8 & -32:46:43 & 15.01 & 16 & 13410 &  26 & 6 & s8 \nl
046 & B133000.5-342742\tablenotemark{b} & 13:30:00.4 & -34:27:42 & 15.67 &  5 & 36879 &  77 & 0 & DPP \nl
047 &                 & 13:30:00.7 & -32:46:20 & 15.89 &  5 & 13451 &  51 & 2 & s8 \nl
048 &                 & 13:30:02.3 & -34:43:34 & 16.00 &  5 &   --  &  -- & -- & -- \nl
049 &                 & 13:30:02.8 & -34:26:34 & 16.32 &  5 &   --  &  -- & -- & -- \nl
050 &                 & 13:30:03.8 & -33:03:46 & 16.31 &  5 &   --  &  -- & -- & -- \nl
051 & QRM~1330-32~04\tablenotemark{e},~B133004.4-325449\tablenotemark{b} & 13:30:04.3 & -32:54:50 & 15.04 & -2 & 15749 &  26 & 0 & s5 \nl
052 &                 & 13:30:05.9 & -33:07:02 & 16.09 &  5 &   --  &  -- & -- & -- \nl
053 &                 & 13:30:07.9 & -32:51:31 & 15.66 &  5 & 15127 &  30 & 0 & s8 \nl
054 &                 & 13:30:09.6 & -32:49:32 & 15.95 &  5 &   --  &  -- & -- & -- \nl
055 & B133010.9-325038\tablenotemark{b} & 13:30:11.3 & -32:50:38 & 14.50 & -2 & 15006 &  25 & 0 & s8 \nl
056 &                 & 13:30:11.7 & -34:02:28 & 15.08 &  5 & 14435 &  37 & 3 & s5 \nl
057 & QRM~1330-32~08\tablenotemark{e}          & 13:30:13.0 & -32:57:32 & 14.88 &  5 & 13732 &  40 & 0 & s5 \nl
058 &                 & 13:30:13.6 & -32:54:47 & 16.10 &  5 &   --  &  -- & -- & -- \nl
059 & B133014.9-330241\tablenotemark{b} & 13:30:14.9 & -33:02:41 & 15.53 &  5 & 14838 & 115 & 0 & DPP \nl
060 & B133015.5-341136\tablenotemark{b} & 13:30:15.5 & -34:11:36 & 15.04 &  5 &  7451 &  38 & 6 & s5 \nl
061 &                 & 13:30:24.7 & -32:57:45 & 16.25 & -2 &   --  &  -- & -- & -- \nl
241 & 383~G~17          & 13:30:25.1 & -34:12:38 & 15.70 & 22 &  3471 &  10 & 0 & CFCQ \nl
062 &                 & 13:30:30.3 & -33:41:15 & 15.70 &  5 &   --  &  -- & -- & -- \nl
063 &                 & 13:30:31.0 & -32:49:35 & 15.33 &  5 & 14490 &  33 & 0 & s10 \nl
064 &                 & 13:30:32.0 & -32:59:34 & 15.90 & -2 &   --  &  -- & -- & -- \nl
065 & B133034.1-334134\tablenotemark{b} & 13:30:34.0 & -33:41:35 & 16.28 &  5 & 22010 & 143 & 0 & DPP \nl
066 &                 & 13:30:34.5 & -33:40:49 & 15.66 &  5 &   --  &  -- & -- & -- \nl
067 & 383~G~18          & 13:30:34.7 & -33:45:33 & 14.72 &  5 &  3721 &  17 & 3 & s1 \nl
068 &                 & 13:30:45.8 & -32:57:04 & 15.68 &  5 &   --  &  -- & -- & -- \nl
069 &                 & 13:30:51.9 & -33:08:41 & 15.90 &  5 &   --  &  -- & -- & -- \nl
070 &                 & 13:30:55.0 & -32:54:46 & 15.10 &  5 & 15355 &  27 & 0 & s10 \nl
071 &                 & 13:30:55.9 & -32:57:00 & 16.12 &  5 & 14957 &  37 & 0 & s11 \nl
072 &                 & 13:30:56.0 & -32:54:34 & 15.10 & 33 &   --  &  -- & -- & -- \nl
073 &                 & 13:30:57.9 & -32:58:06 & 15.94 & 33 &   --  &  -- & -- & -- \nl
074 &                 & 13:30:58.8 & -33:21:14 & 16.22 &  5 &   --  &  -- & -- & -- \nl
075 &                 & 13:31:00.5 & -33:03:53 & 15.15 &  5 & 14595 &  78 & 0 & s1 \nl
076 &                 & 13:31:02.0 & -32:45:29 & 14.39 & -2 & 13361 &  45 & 0 & s1 \nl
077 &                 & 13:31:02.8 & -33:06:25 & 15.26 &  5 & 14999 &  41 & 0 & s5 \nl
078 &                 & 13:31:03.3 & -34:38:53 & 15.24 &  5 &  8177 &  83 & 7 & s5 \nl
079 &                 & 13:31:09.2 & -33:32:19 & 16.07 &  5 &   --  &  -- & -- & -- \nl
080 &                 & 13:31:13.4 & -33:15:07 & 16.04 &  5 &   --  &  -- & -- & -- \nl
081 &                 & 13:31:17.1 & -32:59:53 & 15.25 &  5 & 13762 &  55 & 5 & s10 \nl
082 &                 & 13:31:20.4 & -33:16:13 & 15.30 &  5 & 14031 &  23 & 5 & s10 \nl
083 & 383~G~22,~B133121.8-324317\tablenotemark{b} & 13:31:21.9 & -32:43:17 & 14.38 &  0 & 15207 &  35 & 0 & s1 \nl
084 &                 & 13:31:26.9 & -33:48:54 & 13.97 &  5 &  3692 &  29 & 0 & s3 \nl
085 &                 & 13:31:30.4 & -32:58:21 & 16.10 &  5 &   --  &  -- & -- & -- \nl
086 &                 & 13:31:31.6 & -33:42:35 & 16.28 &  5 &   --  &  -- & -- & -- \nl
087 &                 & 13:31:35.9 & -34:25:02 & 15.22 &  5 & 14848 &  49 & 0 & s10 \nl
088 &                 & 13:31:39.9 & -32:43:02 & 15.48 & -2 & 14946 &  34 & 0 & s11 \nl
089 &                 & 13:31:46.6 & -32:51:59 & 15.61 &  5 &   --  &  -- & -- & -- \nl
090 &                 & 13:31:48.0 & -33:12:07 & 14.47 &  5 & 13866 &  40 & 2 & s2 \nl
091 & B133148.9-344649\tablenotemark{b} & 13:31:48.6 & -34:46:43 & 15.70 &  5 &  2367 &  21 & 0 & DPP \nl
092 & 383~G~24          & 13:31:50.1 & -33:15:48 & 15.08 &  5 &  3270 &  79 & 8 & s8 \nl
093 &                 & 13:31:50.2 & -33:46:55 & 16.21 &  5 &   --  &  -- & -- & -- \nl
094 & MCG-06-30-010,~383~G~25 & 13:31:54.3 & -34:03:17 & 13.22 &  0 &  3994 &  35 & 1 & s1 \nl
095 &                 & 13:31:56.7 & -33:19:40 & 16.14 &  5 &   --  &  -- & -- & -- \nl
096 &                 & 13:31:57.0 & -32:53:30 & 14.51 &  5 & 15452 &  36 & 2 & s5 \nl
097 &                 & 13:31:57.9 & -33:10:44 & 16.12 &  5 &   --  &  -- & -- & -- \nl
098 &                 & 13:31:58.7 & -32:49:24 & 15.34 &  5 & 15149 &  30 & 0 & s10 \nl
099 & CSRG~0728\tablenotemark{f}       & 13:32:02.3 & -33:37:20 & 14.28 &  0 &  3809 &  71 & 0 & s1 \nl
100 &                 & 13:32:02.5 & -32:46:53 & 16.00 &  5 &   --  &  -- & -- & -- \nl
101 &                 & 13:32:04.3 & -33:49:39 & 15.84 &  5 &   --  &  -- & -- & -- \nl
102 &                 & 13:32:05.8 & -33:57:39 & 15.56 &  5 &   --  &  -- & -- & -- \nl
103 &                 & 13:32:08.3 & -33:40:36 & 15.47 &  5 &  3949 &  65 & 0 & s5 \nl
104 & NGC~5215A,~383~G~29A & 13:32:15.9 & -33:13:29 & 13.27 & 16 &  3838 &  27 & 0 & DC1 \nl
105 &                 & 13:32:16.1 & -34:01:46 & 15.71 &  5 &   --  &  -- & -- & -- \nl
106 &                 & 13:32:16.8 & -34:46:55 & 15.97 &  5 &   --  &  -- & -- & -- \nl
107 &                 & 13:32:17.0 & -34:02:01 & 15.71 & 16 &   --  &  -- & -- & -- \nl
108 & B133217.4-334646\tablenotemark{b} & 13:32:17.1 & -33:46:46 & 14.69 &  5 &  3950 &  40 & 0 & s8 \nl
109 & NGC~5215B,~383~G~29 & 13:32:18.6 & -33:13:39 & 12.86 & 16 &  4013 &  25 & 0 & DC1 \nl
111 &                 & 13:32:25.3 & -33:49:19 & 15.06 &  5 & 13378 &  53 & 5 & s8 \nl
112 & MCG-05-32-043,~383~G~30 & 13:32:25.9 & -33:38:34 & 12.91 &  5 &  3617 &  34 & 5 & s11 \nl
113 & MCG-06-30-013,~383~G~31 & 13:32:29.8 & -33:57:02 & 12.85 &  5 &  7130 &  15 & 2 & DC3 \nl
114 & 383~G~32,~B133231.3-335420\tablenotemark{b} & 13:32:31.3 & -33:54:20 & 13.79 &  5 &  7465 &  93 & 0 & s2 \nl
115 &                 & 13:32:33.5 & -33:37:54 & 16.36 &  5 &   --  &  -- & -- & -- \nl
116 & MCG-06-30-014,~383~G~33 & 13:32:34.6 & -33:56:06 & 14.49 &  5 &  7271 &  34 & 8 & s5 \nl
117 &                 & 13:32:35.9 & -33:17:34 & 15.48 &  5 & 14294 &  29 & 0 & s11 \nl
118 &                 & 13:32:40.0 & -33:58:27 & 15.97 &  5 &   --  &  -- & -- & -- \nl
119 &                 & 13:32:44.0 & -34:15:33 & 15.66 &  5 &   --  &  -- & -- & -- \nl
245 &                 & 13:32:45.0 & -33:15:52 & 16.64 &  5 &  7561 &  41 & 3 & s11 \nl
120 &                 & 13:32:49.9 & -32:39:12 & 15.77 &  5 &   --  &  -- & -- & -- \nl
121 &                 & 13:32:58.0 & -33:59:14 & 14.77 &  5 &  3309 &  30 & 0 & s8 \nl
122 &                 & 13:33:01.9 & -34:23:56 & 15.71 &  5 &   --  &  -- & -- & -- \nl
123 & MCG-06-30-015,~383~G~35 & 13:33:01.9 & -34:02:26 & 12.69 &  5 &  2358 &  19 & 3 & s6 \nl
124 & NGC~5220,~383~G~36 & 13:33:05.6 & -33:11:51 & 12.06 & -2 &  4213 &  20 & 0 & s6 \nl
125 &                 & 13:33:10.5 & -32:45:08 & 15.95 &  5 &   --  &  -- & -- & -- \nl
126 & MCG-05-32-048,~383~G~37 & 13:33:14.3 & -32:45:21 & 13.23 &  5 &  3563 &  31 & 3 & DC2 \nl
127 &                 & 13:33:19.6 & -33:38:22 & 15.37 & -2 &  3542 &  43 & 0 & s10 \nl
128 &                 & 13:33:21.9 & -34:07:11 & 15.13 &  5 &  3829 &  30 & 0 & s10 \nl
129 &                 & 13:33:23.1 & -33:15:47 & 16.11 &  5 &   --  &  -- & -- & -- \nl
130 &                 & 13:33:24.2 & -33:21:49 & 14.74 &  5 &  3406 &  78 & 0 & s8 \nl
131 & 383~G~38          & 13:33:27.3 & -32:58:17 & 13.64 &  5 &  7507 &  50 & 0 & s1 \nl
132 &                 & 13:33:34.8 & -34:00:56 & 15.53 &  5 &   --  &  -- & -- & -- \nl
133 &                 & 13:33:36.7 & -33:25:36 & 16.34 &  5 &   --  &  -- & -- & -- \nl
134 & B133337.8-325015\tablenotemark{b} & 13:33:37.8 & -32:50:16 & 15.53 &  5 & 15688 & 105 & 0 & DPP \nl
135 & B133341.2-334911\tablenotemark{b} & 13:33:41.1 & -33:49:11 & 13.85 &  5 &  3850 &  20 & 2 & s8 \nl
136 &                 & 13:33:44.9 & -33:39:30 & 16.34 & -2 &   --  &  -- & -- & -- \nl
137 & IC~4296,~383~G~39 & 13:33:47.1 & -33:42:40 & 10.33 & -5 &  3785 &  19 & 0 & s6 \nl
138 &                 & 13:33:47.5 & -33:51:02 & 16.02 &  5 &   --  &  -- & -- & -- \nl
139 &                 & 13:33:50.5 & -32:53:34 & 15.49 & 33 &  7166 &  34 & 8 & s11 \nl
140 &                 & 13:33:52.6 & -34:00:24 & 16.34 &  5 &   --  &  -- & -- & -- \nl
141 & IC~4299,~383~G~42 & 13:33:55.6 & -33:48:41 & 11.99 &  5 &  4045 &  52 & 0 & s1 \nl
142 &                 & 13:33:56.4 & -33:28:30 & 16.18 &  5 &   --  &  -- & -- & -- \nl
143 &                 & 13:34:02.7 & -33:32:33 & 15.88 &  5 &   --  &  -- & -- & -- \nl
144 & B133405.6-340842\tablenotemark{b} & 13:34:05.6 & -34:08:43 & 15.29 &  5 &  4109 &  23 & 4 & s10 \nl
145 & AM~1334-333\tablenotemark{c}     & 13:34:06.2 & -33:29:49 & 13.78 &  5 & 11567 &  24 & 0 & s6 \nl
146 &                 & 13:34:06.8 & -33:34:15 & 16.31 &  5 &   --  &  -- & -- & -- \nl
147 & AM~1334-333\tablenotemark{c}     & 13:34:07.1 & -33:29:27 & 14.90 &  5 &  4099 &  47 & 0 & s6 \nl
148 &                 & 13:34:09.4 & -33:55:01 & 15.01 &  5 & 21348 &  35 & 0 & s8 \nl
149 & CSRG~0731\tablenotemark{f},~B133411.1-333408\tablenotemark{b} & 13:34:10.9 & -33:34:04 & 14.07 &  5 & 14091 &  35 & 2 & s6 \nl
150 &                 & 13:34:17.0 & -33:30:29 & 15.71 &  5 &   --  &  -- & -- & -- \nl
151 & J133709.39-334731.9\tablenotemark{g} & 13:34:17.0 & -33:32:16 & 16.35 &  5 &   --  &  -- & -- & -- \nl
152 & B133428.4-343809\tablenotemark{b} & 13:34:28.3 & -34:38:06 & 15.83 &  5 & 29247 & 103 & 0 & DPP \nl
153 &                 & 13:34:29.1 & -33:29:07 & 16.10 & -2 &   --  &  -- & -- & -- \nl
154 & MCG-05-32-052,~383~G~44 & 13:34:36.6 & -32:45:10 & 13.29 &  5 &  3776 &  16 & 3 & s3 \nl
155 &                 & 13:34:38.4 & -33:13:23 & 15.41 &  5 & 11275 &  26 & 8 & s10 \nl
156 &                 & 13:34:39.8 & -34:24:18 & 15.20 &  5 & 17560 &  38 & 6 & s10 \nl
157 &                 & 13:34:42.4 & -34:01:31 & 16.30 &  5 &   --  &  -- & -- & -- \nl
158 & B133445.1-341642\tablenotemark{b} & 13:34:45.1 & -34:16:42 & 15.74 & -2 & 22108 &  68 & 0 & DPP \nl
159 & MCG-05-32-053,~383~G~45 & 13:34:47.6 & -33:33:22 & 12.56 & -2 &  3914 &  19 & 0 & s6 \nl
160 & B133449.6-333555\tablenotemark{b} & 13:34:49.7 & -33:35:53 & 15.19 &  5 & 11292 &  62 & 4 & s8 \nl
161 &                 & 13:34:55.0 & -32:39:20 & 15.10 &  5 &  7447 &  24 & 0 & s10 \nl
162 & 383~G~46          & 13:34:55.3 & -34:33:14 & 15.04 &  5 & 13014 &  57 & 5 & s5 \nl
163 &                 & 13:34:55.9 & -34:14:19 & 16.28 &  5 &   --  &  -- & -- & -- \nl
164 &                 & 13:34:57.3 & -33:34:25 & 16.08 &  5 &   --  &  -- & -- & -- \nl
165 &                 & 13:35:01.1 & -33:33:21 & 14.15 &  5 & 11402 &  28 & 0 & s7 \nl
166 &                 & 13:35:02.0 & -34:28:10 & 16.17 &  5 &   --  &  -- & -- & -- \nl
167 &                 & 13:35:08.7 & -33:43:59 & 15.82 & -2 & 37229 &  51 & 0 & s8 \nl
168 &                 & 13:35:08.8 & -33:44:19 & 14.66 &  5 &  3698 &  29 & 0 & s8 \nl
169 & MCG-05-32-054,~383~G~48 & 13:35:09.5 & -33:15:52 & 13.28 &  5 &  3702 &  25 & 3 & s4 \nl
170 &                 & 13:35:11.7 & -33:25:48 & 16.02 &  5 &   --  &  -- & -- & -- \nl
171 & MCG-05-32-055,~383~G~49 & 13:35:11.7 & -33:37:09 & 12.79 & -2 &  3925 &  27 & 0 & s11 \nl
172 &                 & 13:35:13.4 & -34:08:33 & 16.12 &  5 &   --  &  -- & -- & -- \nl
173 & B133515.2-344829\tablenotemark{b} & 13:35:14.8 & -34:48:20 & 15.67 &  5 & 15162 &  92 & 0 & DPP \nl
174 &                 & 13:35:15.1 & -34:08:06 & 15.58 &  5 &   --  &  -- & -- & -- \nl
175 &                 & 13:35:15.3 & -33:51:47 & 15.51 &  5 &   --  &  -- & -- & -- \nl
176 &                 & 13:35:16.9 & -33:39:22 & 16.34 & -2 &   --  &  -- & -- & -- \nl
177 &                 & 13:35:21.5 & -33:29:11 & 16.06 &  5 &   --  &  -- & -- & -- \nl
178 &                 & 13:35:29.4 & -34:36:19 & 16.13 &  5 &   --  &  -- & -- & -- \nl
179 &                 & 13:35:32.6 & -34:35:06 & 16.19 &  5 &   --  &  -- & -- & -- \nl
180 &                 & 13:35:34.6 & -34:21:28 & 16.30 & -3 &   --  &  -- & -- & -- \nl
181 &                 & 13:35:37.6 & -32:52:04 & 16.19 &  5 &   --  &  -- & -- & -- \nl
182 &                 & 13:35:46.9 & -33:05:54 & 14.75 &  5 & 12731 &  15 & 0 & s7 \nl
183 &                 & 13:35:48.4 & -33:15:50 & 14.94 &  5 & 22237 &  66 & 0 & s2 \nl
184 &                 & 13:35:48.6 & -33:13:12 & 16.30 &  5 &   --  &  -- & -- & -- \nl
185 &                 & 13:36:02.8 & -32:38:24 & 16.29 &  5 &   --  &  -- & -- & -- \nl
186 &                 & 13:36:09.1 & -32:38:25 & 16.10 & -2 &   --  &  -- & -- & -- \nl
187 &                 & 13:36:11.0 & -32:39:17 & 16.16 & -2 &   --  &  -- & -- & -- \nl
188 &                 & 13:36:21.1 & -32:57:09 & 15.80 &  5 &   --  &  -- & -- & -- \nl
189 &                 & 13:36:25.2 & -33:09:51 & 15.15 &  5 & 12660 &  53 & 8 & s10 \nl
190 &                 & 13:36:30.0 & -34:06:25 & 15.61 &  5 &   --  &  -- & -- & -- \nl
191 &                 & 13:36:33.5 & -34:17:58 & 16.33 &  5 &   --  &  -- & -- & -- \nl
242 & HSTGC~07274-00974\tablenotemark{d} & 13:36:34.2 & -34:01:07 & 15.85 &  5 &   --  &  -- & -- & -- \nl
192 & 383~G~55          & 13:36:34.7 & -33:55:59 & 14.09 &  5 &  7691 &  10 & 0 & MFB \nl
193 & B133644.2-334554\tablenotemark{b} & 13:36:43.7 & -33:45:47 & 15.43 &  5 & 15222 &  32 & 0 & s6 \nl
194 & 383~G~56          & 13:36:45.2 & -32:38:21 & 14.75 &  5 & 12260 &  30 & 4 & s5 \nl
195 &                 & 13:36:45.7 & -33:55:19 & 15.66 &  5 &   --  &  -- & -- & -- \nl
196 & B133656.1-332407\tablenotemark{b} & 13:36:56.0 & -33:24:01 & 14.62 &  5 & 15494 &  47 & 0 & s2 \nl
197 & B133656.6-324951\tablenotemark{b} & 13:36:56.6 & -32:49:51 & 15.42 &  5 & 15327 &  15 & 4 & s10 \nl
199 &                 & 13:37:04.7 & -34:04:22 & 15.79 &  5 &   --  &  -- & -- & -- \nl
200 & 383~G~59          & 13:37:11.2 & -32:38:32 & 15.43 &  5 &  7364 &  28 & 6 & s5 \nl
201 & B133712.1-324853\tablenotemark{b} & 13:37:12.1 & -32:48:53 & 15.89 &  5 & 11841 &  59 & 0 & DPP \nl
202 &                 & 13:37:14.6 & -33:50:08 & 16.05 &  5 &   --  &  -- & -- & -- \nl
203 &                 & 13:37:21.4 & -33:40:28 & 14.67 &  5 & 15088 &  31 & 0 & s2 \nl
204 &                 & 13:37:34.6 & -33:44:14 & 15.78 & -2 &   --  &  -- & -- & -- \nl
205 &                 & 13:37:35.9 & -32:42:56 & 15.70 &  5 &   --  &  -- & -- & -- \nl
206 & B133737.8-334408\tablenotemark{b} & 13:37:37.8 & -33:44:08 & 16.04 &  5 & 14867 &  61 & 0 & DPP \nl
207 & MCG-05-32-063,~383~G~60 & 13:37:38.5 & -33:24:09 & 12.81 &  5 &  3890 &  40 & 3 & s1 \nl
208 &                 & 13:37:39.8 & -33:44:22 & 14.95 & -3 & 14928 &  35 & 4 & s7 \nl
209 &                 & 13:37:45.6 & -33:44:17 & 14.32 & -3 & 14840 &  28 & 0 & s7 \nl
210 &                 & 13:37:45.8 & -34:08:18 & 16.20 &  5 &   --  &  -- & -- & -- \nl
211 &                 & 13:37:55.0 & -34:06:26 & 15.82 &  5 &   --  &  -- & -- & -- \nl
212 & B133801.9-332524\tablenotemark{b} & 13:38:02.2 & -33:25:19 & 15.57 &  5 & 15167 &  66 & 0 & DPP \nl
213 &                 & 13:38:03.8 & -33:04:57 & 14.59 &  5 & 12184 &  56 & 7 & s9 \nl
214 &                 & 13:38:07.7 & -33:50:57 & 15.99 &  5 &   --  &  -- & -- & -- \nl
215 &                 & 13:38:08.3 & -33:56:00 & 15.61 &  5 &   --  &  -- & -- & -- \nl
216 &                 & 13:38:10.1 & -34:04:55 & 16.30 &  5 &   --  &  -- & -- & -- \nl
217 & B133812.4-340713\tablenotemark{b} & 13:38:12.9 & -34:07:04 & 16.28 &  5 & 15485 &  98 & 0 & DPP \nl
218 &                 & 13:38:13.9 & -33:16:47 & 16.09 & -2 &   --  &  -- & -- & -- \nl
219 &                 & 13:38:16.3 & -32:47:54 & 15.40 &  5 &  4494 &  44 & 8 & s11 \nl
220 &                 & 13:38:21.9 & -34:35:37 & 16.32 &  5 &   --  &  -- & -- & -- \nl
221 &                 & 13:38:26.4 & -34:07:16 & 15.40 & -2 & 16005 &  27 & 0 & s10 \nl
222 &                 & 13:38:28.5 & -33:15:16 & 15.89 &  5 &   --  &  -- & -- & -- \nl
223 &                 & 13:38:30.4 & -34:38:14 & 15.64 & -5 &   --  &  -- & -- & -- \nl
224 &                 & 13:38:34.9 & -34:06:53 & 15.08 &  5 & 16013 &  27 & 0 & s8 \nl
225 &                 & 13:38:39.3 & -33:19:46 & 16.17 &  5 &   --  &  -- & -- & -- \nl
244 &                 & 13:38:39.5 & -33:41:15 & 17.13 & -5 & 15288 &  32 & 0 & s11 \nl
226 & B133839.8-340745\tablenotemark{b} & 13:38:40.4 & -34:07:37 & 15.75 &  5 & 16922 &  36 & 0 & s8 \nl
227 & B133840.8-334114\tablenotemark{b} & 13:38:41.0 & -33:41:14 & 15.49 &  5 & 15425 &  44 & 6 & s11 \nl
228 &                 & 13:38:46.3 & -34:13:12 & 15.65 &  5 &   --  &  -- & -- & -- \nl
230 &                 & 13:38:47.7 & -34:47:10 & 16.21 &  5 &   --  &  -- & -- & -- \nl
229 &                 & 13:38:47.7 & -34:03:40 & 16.34 &  5 &   --  &  -- & -- & -- \nl
231 &                 & 13:38:48.4 & -34:10:44 & 14.49 & -5 & 15620 &  77 & 0 & s1 \nl
232 & B133849.2-334906\tablenotemark{b} & 13:38:49.4 & -33:49:05 & 15.96 &  5 & 14610 &  47 & 0 & DPP \nl
233 &                 & 13:38:52.9 & -33:57:47 & 15.56 &  5 &   --  &  -- & -- & -- \nl
234 & B133857.2-325558\tablenotemark{b} & 13:38:57.2 & -32:56:02 & 15.63 &  5 & 11945 &  65 & 0 & DPP \nl
235 &                 & 13:39:01.1 & -33:50:50 & 15.59 & -2 &   --  &  -- & -- & -- \nl
236 &                 & 13:39:05.1 & -33:19:43 & 15.65 &  5 &   --  &  -- & -- & -- \nl
238 &                 & 13:39:06.2 & -33:48:14 & 16.10 & -2 &   --  &  -- & -- & -- \nl
237 &                 & 13:39:06.2 & -33:43:54 & 16.30 &  5 &   --  &  -- & -- & -- \nl
239 &                 & 13:39:08.2 & -33:05:30 & 16.02 &  5 &   --  &  -- & -- & -- \nl
240 & 383~G~63          & 13:39:10.6 & -34:17:18 & 14.72 &  5 & 14547 &  28 & 7 & s8 \nl
\enddata
\tablenotetext{a}{Identification in  Gregorini et al. (1994).}
\tablenotetext{b}{Identification in Drinkwater et al. (1999).}
\tablenotetext{c}{Identification in Arp \& Madore (1987).}
\tablenotetext{d}{Identification in Lasker et al. (1990).}
\tablenotetext{e}{Identification in Quintana et al. (1995).}
\tablenotetext{f}{Identification in Buta (1995).}
\tablenotetext{g}{Galaxy identified by the Mount Stromlo Abell Cluster Supernova Search Team (1997)}
\tablerefs{
CFCQ: Cote et al. (1997) ;
DC1: da Costa et al. (1986) ; DC2: da Costa et al. (1987);
DC3: da Costa (1992), private communication; DPP: Drinkwater et al. (1999);
MFB: Matthewson et al. (1992);QRM: Quintana et al. (1995).;
T98: Theureau et al. (1998)}
\end{deluxetable}

\begin{deluxetable}{lccc}
\small
\tablewidth{0pc}
\tablecaption{Dynamical Parameters of Clusters}
\tablehead{
\colhead {Parameter} & \colhead {Units}  & \colhead{A3565}      & \colhead {A3560}
}
\startdata
$\alpha$ (1950.0) & {} &  13$^h$~33$^m$~04$^s~\pm 4^s$&
13$^h$~30$^m$~12$^s~\pm 2^s$ \nl
$\delta$ (1950.0) & {} & -33\deg 31$'$ 50$''~\pm49''$ &-32\deg 53$'$ 26$''~\pm20''$\nl
$l$               & {} & 313.41\deg & 312.87\deg \nl
$b$               & {}  & 28.18\deg &  28.92\deg \nl
N$_g$             & {}           & 29        & 32 \nl
$v_\odot$         & \kms         & 3759 $\pm$ 9    &14645 $\pm$ 20 \nl
$v_0$             & \kms         & 3567 $\pm$ 9    &14452 $\pm$ 20 \nl
$\sigma$          & \kms         & 228 (+38, -26)  & 588 (+92, -63)\nl
v$_{bi-weight}$   & \kms         & 3586 $\pm$ 45   & 14470 $\pm$ 123\nl
$\sigma_{bi-weight}$ & \kms      &  236 $\pm$ 69 &  614 $\pm$ 68\nl
R$_h$             & \h1 Mpc      & 0.37$\pm$0.01   & 0.34$\pm$0.09 \nl
R$_p$             & \h1 Mpc      & 0.53$\pm$0.03   & 1.03$\pm$0.04 \nl
t$_c$             & Hubble times  & 0.04       & 0.02 \nl
M$_{VT}$          & \h1 M$_\odot$  & 2.72 (+1.01, -0.71) $\times 10^{13}$ & 1.64 (+0.93, -0.77) $\times$ 10$^{14}$ \nl
M$_p$             & \h1 M$_\odot$  & 3.29 $\pm$ 0.17  $\times$10$^{13}$& 4.90 $\pm$ 0.19 $\times$10$^{14}$    \nl
(M/L)             & (M/L)$_\odot$ & 143           &  760         \nl
\enddata
\end{deluxetable}


\begin{deluxetable}{llrrrr}
\tiny
\tablewidth{0pc}
\tablecaption{Additional radial velocities}
\tablehead{
\colhead {Gal Id.} & \colhead{Other Id.} &\colhead {R.A.} & \colhead {Dec.}  &
\colhead {$v_\odot$} & \colhead {$\pm$} \nl
{}  & {} &  B1950.0       &   B1950.0   &  \kms  & \kms \nl
~~~(1) & ~~~(2) &(3)~~~~  & (4)~~~ & (5)~~~~  & (6)
}
\startdata
HSTGC~07272-01686   &  {}      & 13:21:56.0  &  -34:02:46  &    15074  &     25   \nl    
HSTGC~07268-00028   &  {}      & 13:22:17.0  &  -33:05:27  &    13783  &     26   \nl    
HSTGC~07272-00317   & 382~G~60 & 13:22:37.0  &  -33:32:12  &     8004  &     22   \nl    
HSTGC~07268-01809   & MGP~1133\tablenotemark{a} & 13:23:59.0  &  -31:44:14  &    14008  &     51   \nl    
HSTGC~07268-01860   & MGP~1145\tablenotemark{a}& 13:24:03.0  &  -31:44:56  &    14792  &     31   \nl    
HSTGC~07272-00553   &  {}      & 13:24:58.0  &  -33:44:35  &    15023  &     26   \nl    
HSTGC~07268-02068   & 383~G~01 & 13:25:14.0  &  -33:25:52  &     8091  &     16   \nl    
HSTGC~07272-00677   &  {}      & 13:25:19.0  &  -33:46:53  &    14426  &     26   \nl    
HSTGC~07268-01590   & MGP~2691\tablenotemark{a} & 13:27:06.0  &  -31:53:01  &     3886  &     40   \nl    
HSTGC~07265-01949   & 444~G~60 & 13:27:43.0  &  -31:19:06  &    14489  &     29   \nl    
HSTGC~07269-00325   & MGP~2953\tablenotemark{a} & 13:27:44.0  &  -32:00:24  &     4044  &     20   \nl    
HSTGC~07277-01636   &  {}      & 13:28:11.0  &  -35:54:12  &     7218  &     34   \nl    
HSTGC~07269-00374   & MGP~3249\tablenotemark{a} & 13:28:30.0  &  -31:48:39  &    12956  &     23   \nl    
WMMA~248\tablenotemark{b}  &  {}      & 13:28:44.8  &  -32:45:13  &    13697  &     57   \nl    
HSTGC~07269-01680   &  {}      & 13:28:45.0  &  -32:43:27  &     3579
&     31   \nl    
HSTGC~07269-01750   & AM~1328-325\tablenotemark{c}  & 13:28:49.0  &  -32:59:08  &    16579  &     35   \nl    
HSTGC~07269-01424   & 383~G~12 & 13:28:52.0  &  -33:07:23  &     7738  &     21   \nl    
HSTGC~07269-00860   &  {}      & 13:28:56.0  &  -32:38:15  &    14780  &     48   \nl    
HSTGC~07269-01083NW &  {}      & 13:29:23.6  &  -32:34:30  &    14017  &     24   \nl    
HSTGC~07269-01083SE &  {}      & 13:29:24.0  &  -32:34:33  &    14366  &     25   \nl    
HSTGC~07269-00070   & 444~G~70 & 13:29:56.0  &  -31:42:35  &    11106  &     31   \nl    
HSTGC~07273-01687   & IRAS~13321-3514\tablenotemark{d} & 13:32:06.0  &  -35:14:35  &    15840  &     29   \nl    
WMMA~246\tablenotemark{e}        &  {}      & 13:32:45.0  &  -33:16:34  &    65529  &     71   \nl
HSTGC~07269-01344   &  {}      & 13:33:05.0  &  -32:30:14  &     7483  &     38   \nl    
HSTGC~07273-01829   &  {}      & 13:33:28.0  &  -34:51:28  &     4267  &     30   \nl    
WMMA~247\tablenotemark{f}        &  {}      & 13:34:06.3  &  -33:29:36  &    11479  &     49   \nl
HSTGC~07269-00876   & CSRG~0732\tablenotemark{g} & 13:34:21.0  &  -32:12:21  &    13660  &     50   \nl    
HSTGC~07265-02190   &  {}      & 13:34:48.0  &  -31:31:20  &    11831  &     28   \nl    
HSTGC~07273-00119   & AM~1334-351\tablenotemark{c} & 13:34:55.0  &  -35:14:50  &    15268  &     24   \nl    
HSTGC~07277-00869   & 383~G~47 & 13:34:57.0  &  -35:47:46  &     3629  &     21   \nl    
HSTGC~07266-00111   & CSRG~0737\tablenotemark{g} & 13:36:30.0  &  -31:25:26  &    11574  &     25   \nl    
HSTGC~07266-00139   & IRAS~13365-3116\tablenotemark{d} & 13:36:31.0  &  -31:16:40  &     6932  &     47   \nl    
HSTGC~07270-00593   & 445~G~03 & 13:36:41.0  &  -31:51:03  &     6969  &     43   \nl    
HSTGC~07278-01640   & PL~1\tablenotemark{h}     & 13:37:19.0  &  -35:25:19  &    15638  &     52   \nl    
HSTGC~07270-00628   &  {}      & 13:37:52.0  &  -31:53:58  &    11784  &     25   \nl    
HSTGC~07270-00562   & 445~G~09 & 13:37:57.0  &  -32:24:18  &    11436  &     27   \nl    
HSTGC~07270-00284   &  {}      & 13:38:03.0  &  -33:04:55  &    12155  &    116   \nl    
HSTGC~07270-00724   & 445~G~11 & 13:38:35.0  &  -31:45:01  &     6634  &     60   \nl    
HSTGC~07274-01160   &  {}      & 13:39:49.0  &  -33:44:39  &     5726  &     31   \nl    
HSTGC~07278-01018   & 383~G~62A& 13:39:03.0  &  -35:26:50  &    11467  &     33   \nl    
HSTGC~07278-01224   & 383~G~62 & 13:39:02.4  &  -35:26:35  &     8789  &     37   \nl    
HSTGC~07278-01206   & 383~G~64 & 13:39:13.0  &  -36:05:50  &    11474  &     44   \nl    
HSTGC~07270-00993   &  {}      & 13:40:07.0  &  -31:38:04  &    11639  &     29   \nl    
HSTGC~07274-01486   &  {}      & 13:40:26.0  &  -33:51:32  &    15073  &     40   \nl    
HSTGC~07278-01133   &  {}      & 13:40:41.0  &  -36:09:57  &     4285  &     41   \nl    
HSTGC~07266-00818   &  {}      & 13:42:12.0  &  -31:23:01  &     4678  &     37   \nl    
HSTGC~07270-02135   & 383~G~73 & 13:42:20.0  &  -33:25:44  &    11851  &     26   \nl    
HSTGC~07270-01022   &  {}      & 13:42:24.0  &  -32:23:49  &     9665  &     28   \nl    
HSTGC~07270-02249   &  {}      & 13:43:42.0  &  -33:17:30  &     7521  &     29   \nl    
HSTGC~07270-01038   &  {}      & 13:44:37.0  &  -31:53:31  &    12749  &     37   \nl    
HSTGC~07274-00031   & 383~G~80 & 13:44:50.0  &  -34:43:13  &    11504  &     23   \nl    
HSTGC~07270-01147   & 445~G~33 & 13:45:00.0  &  -32:12:46  &    12229  &     29   \nl    
HSTGC~07283-00985   &  {}      & 13:45:13.0  &  -32:15:36  &    11745  &     29   \nl    
HSTGC~07279-01872   & IRAS~F13454-3132\tablenotemark{d} & 13:45:24.0  &  -31:31:50  &    11428  &     42   \nl    
\enddata
\tablenotetext{a} {Identification in Metcalfe, Godwin \& Peach (1994).}
\tablenotetext{b}{The galaxy wwma~248 is a background object some
2$'$ to the south of HSTGC~07269-01680.}
\tablenotetext{c} {Identification in Arp \& Madore (1987). }
\tablenotetext{d} {Identification in IRAS Faint Source Catalog.}
\tablenotetext{e}{Serendipitous galaxy located between
WMMA~117 and WMMA~245 of Table 4. Coordinates estimated from the 
Digital Sky Survey.}
\tablenotetext{f}{Serendipitous galaxy located between WMMA~145 and
WMMA~147.}
\tablenotetext{g} {Identification in Buta (1995).}
\tablenotetext{h} {Identification in Postman \& Lauer (1995).} 
\end{deluxetable}

\begin{figure}
\plotone{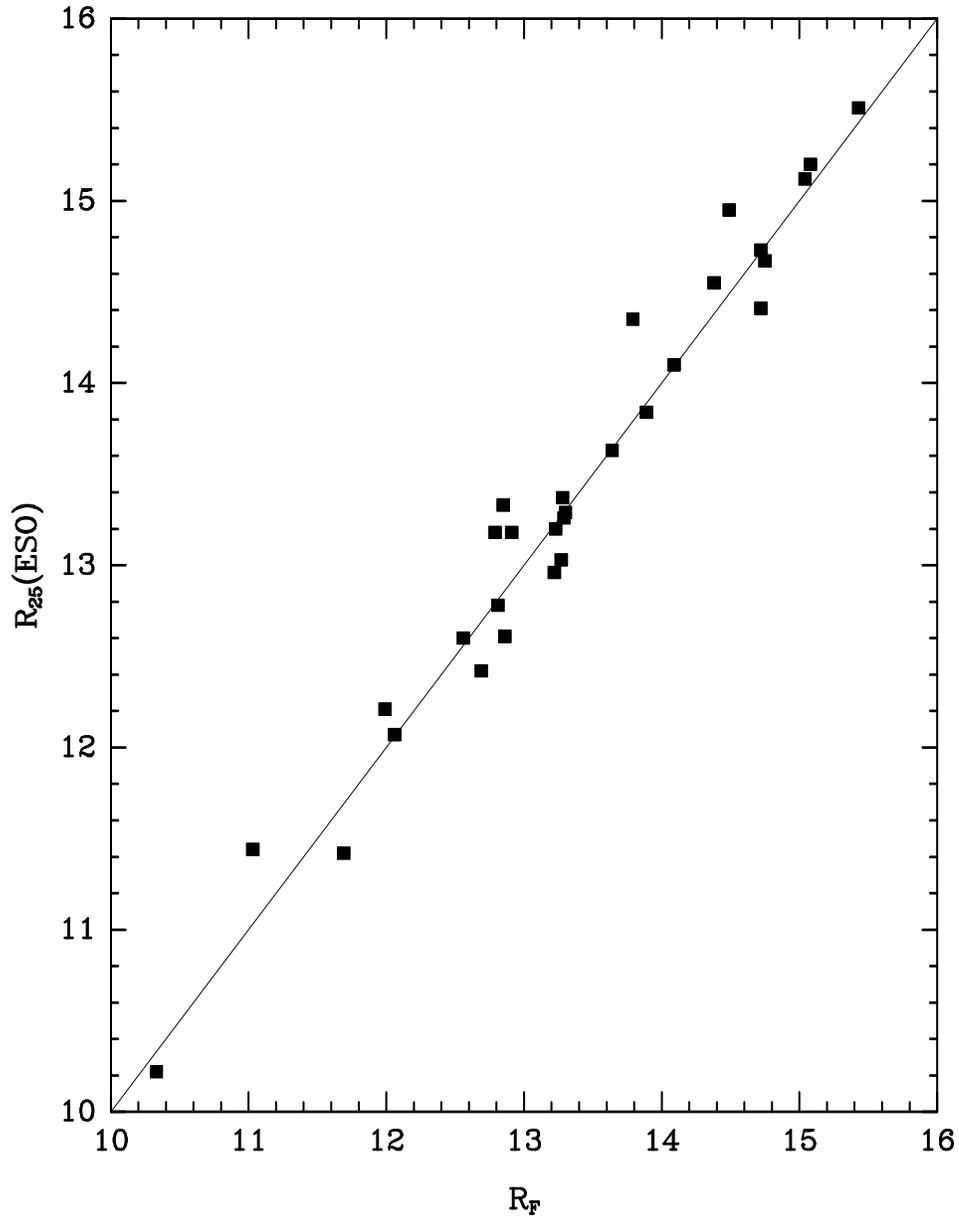}
\caption{
Comparison between our \rf magnitudes and $R_{25}(ESO)$ from ESO--LV. 
}
\end{figure}

\begin{figure}
\plotone{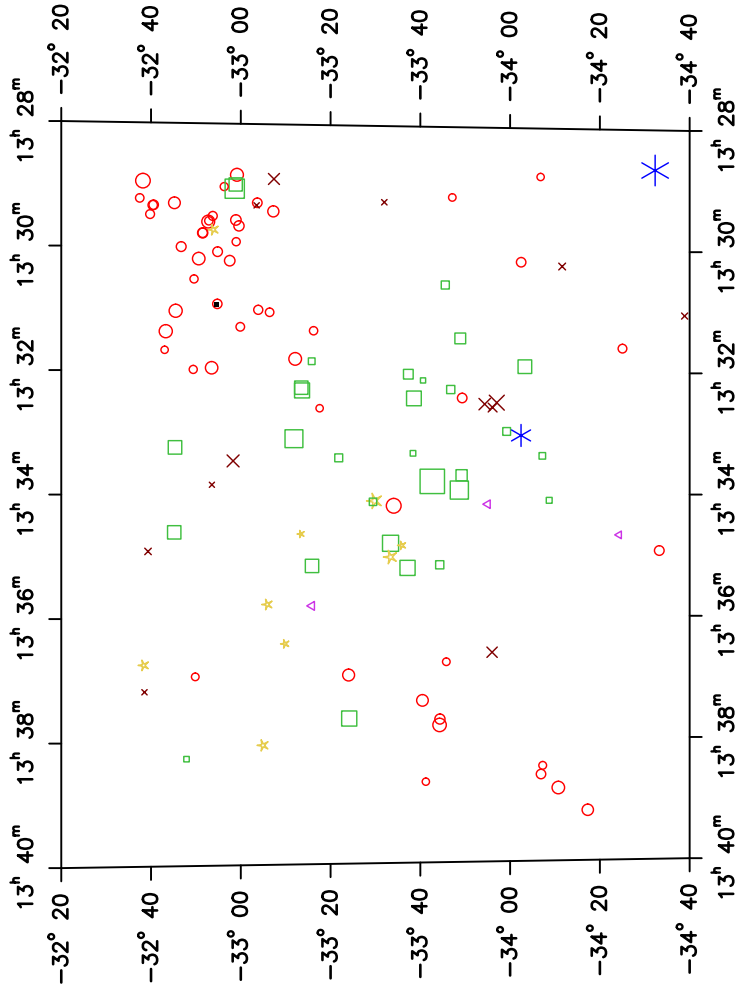}
\caption{
Distribution of galaxies within the surveyed region with
$R_F~ \leq$ 15.50 $mag$. The symbol
sizes represent the apparent magnitude of galaxies. The solid square
represents the galaxy without a redshift measurement, while the
remaining symbols correspond to the following intervals:
asterisks (v $\leq$ 3000 \kms); open squares ( 3000 $<$ v $\leq$
5000 \kms) ; crosses (5000  $<$ v $\leq$ 10000 \kms) ;
stars (10000  $<$ v $\leq$ 13000 \kms); open circles 
(13000  $<$ v $\leq$ 17000 \kms ) and  open triangles 
( v $>$ 17000 \kms).
}
\end{figure}

\begin{figure}
\plotone{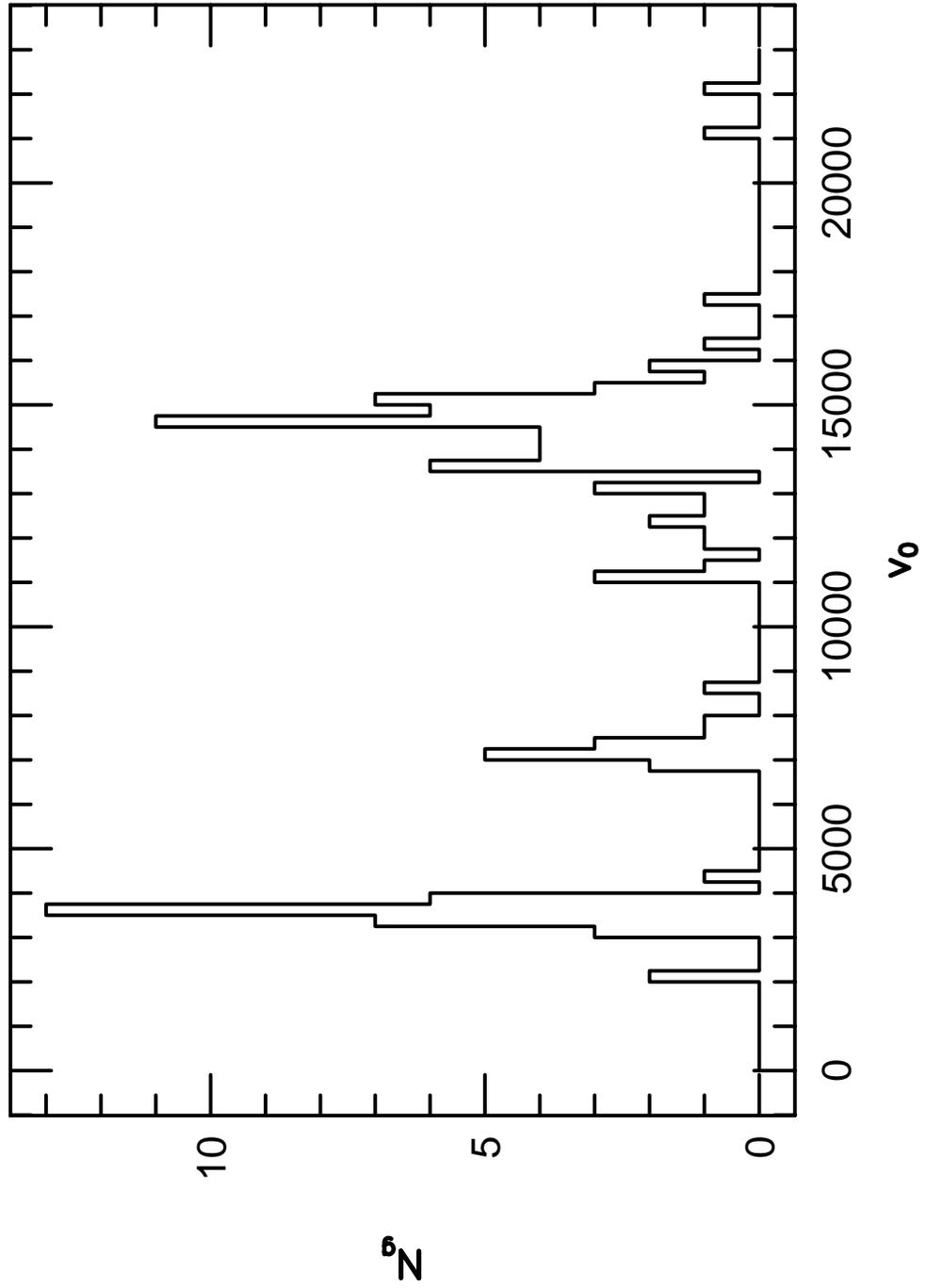}
\caption{
Distribution of radial velocities of galaxies corrected for galactic
rotation within the surveyed region, in bins of 250 \kms.
}
\end{figure}

\begin{figure}
\plotone{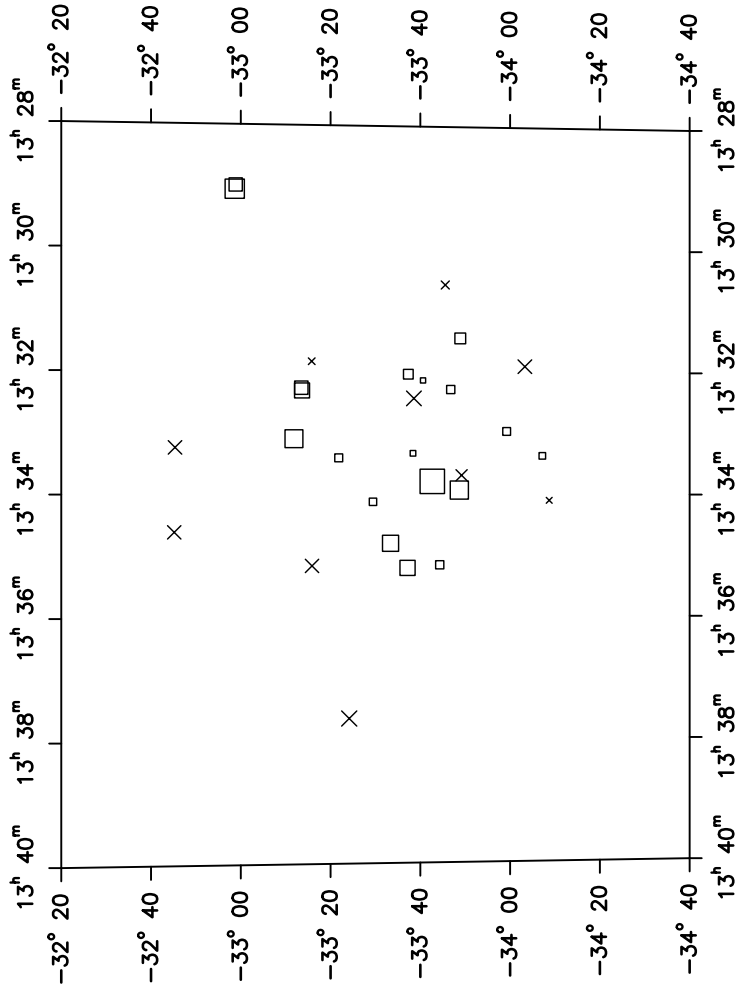}
\caption{
Distribution of galaxies within 3000 and 5000 kms$^{-1}$. Open squares
represent galaxies with absorption line spectra and crosses galaxies
with emission lines. 
}
\end{figure}

\begin{figure}
\plotone{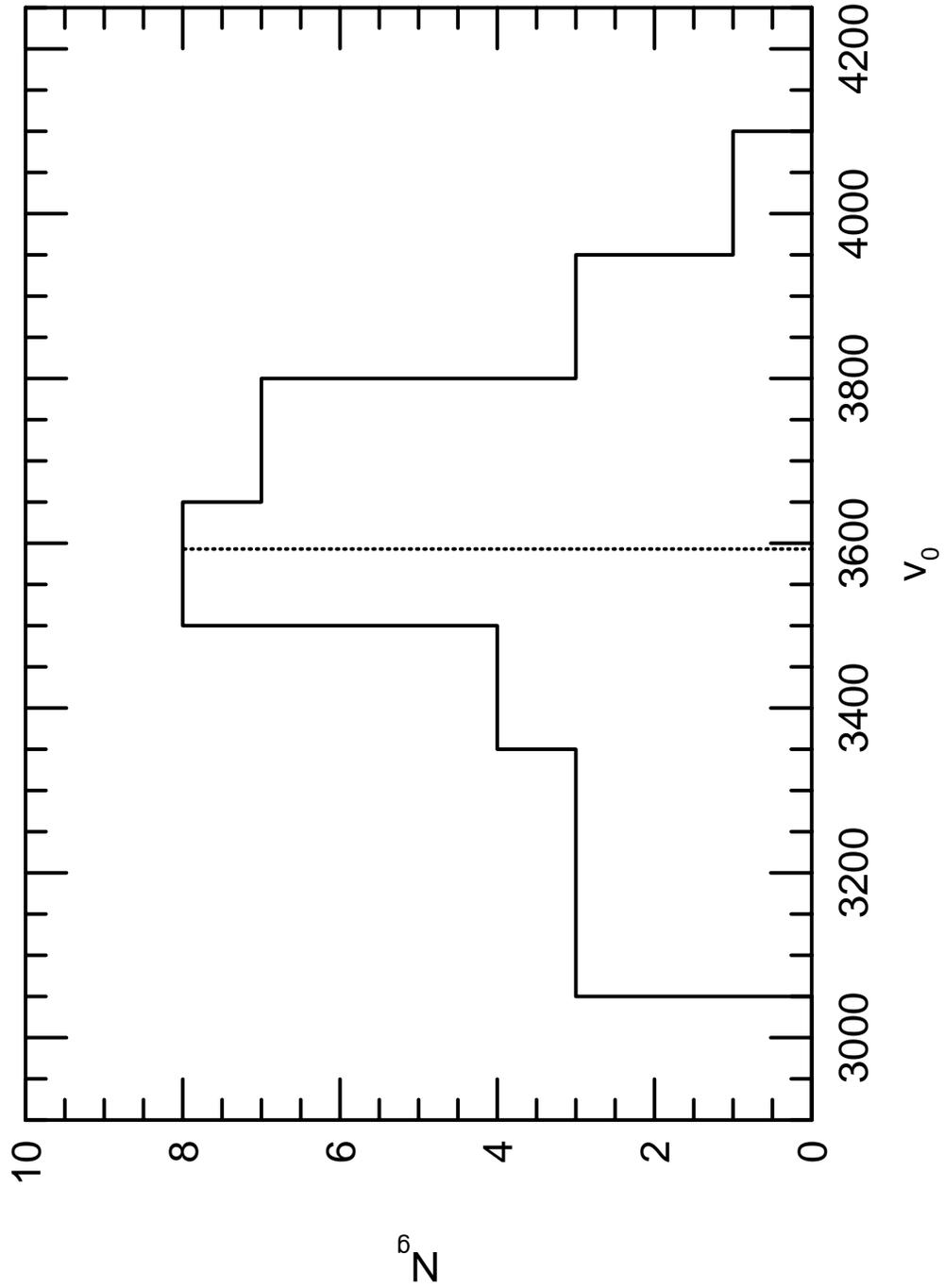}
\caption{
Distribution of radial velocities of galaxies of A3565, corrected
for galactic rotation. The bin size is 150 \kms ; the dotted line represents the radial
velocity of IC~4296.
}
\end{figure}

\begin{figure}
\plotone{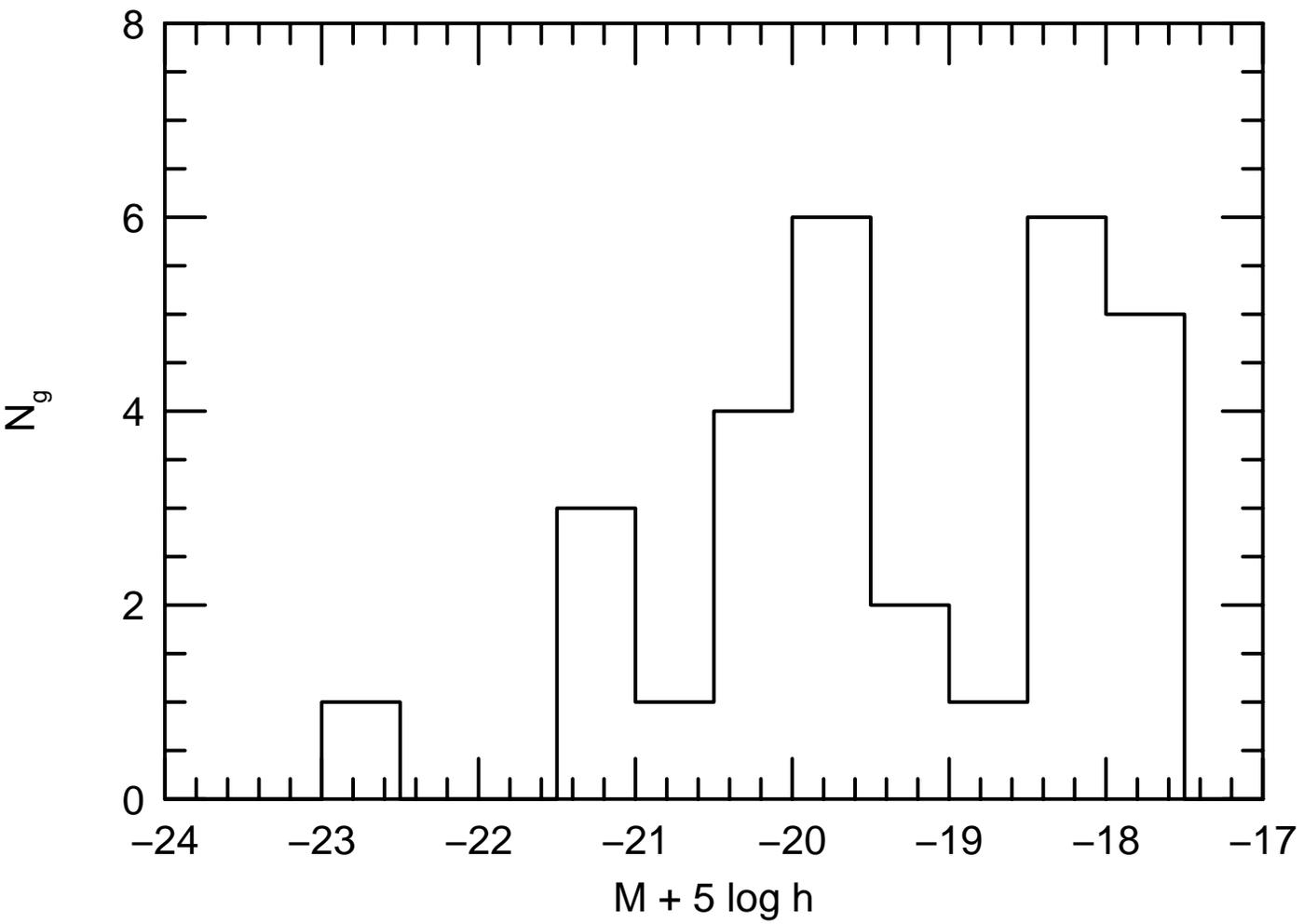}
\caption{
Distribution of absolute magnitudes of A3565 galaxies in 0.5 $mag$
bins. The absolute magnitudes were calculated using the mean group
velocity.
}
\end{figure}

\begin{figure}
\plotone{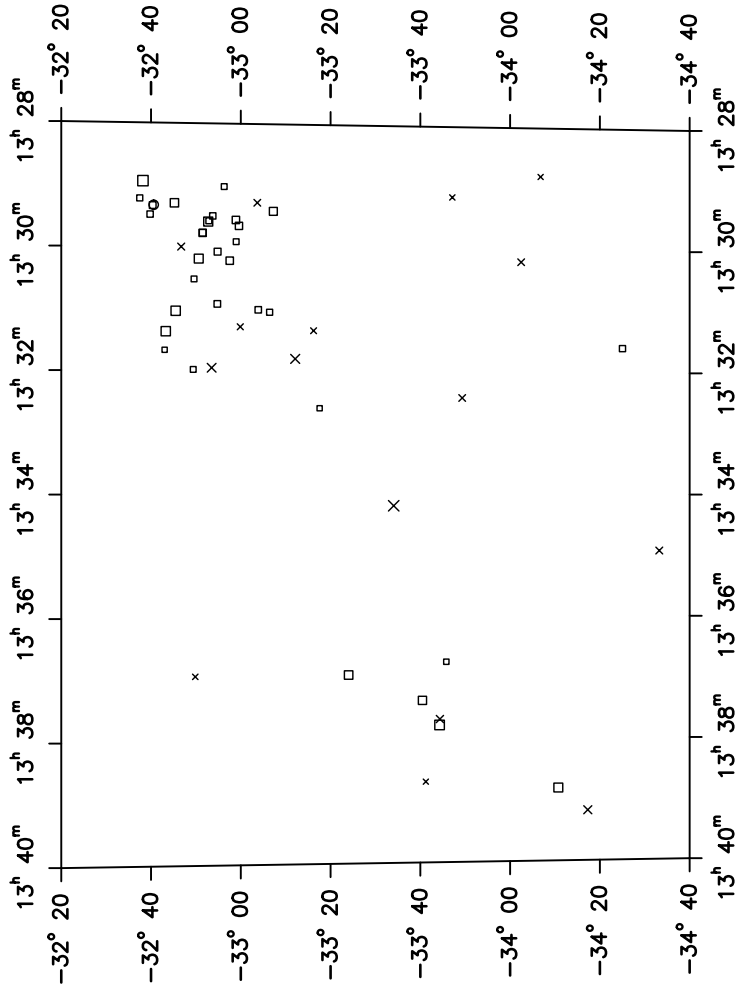}
\caption{
Distribution of galaxies in the radial velocity interval
13000 \kms $\leq v_0 \leq$ 16000 kms$^{-1}$. Galaxies with
absorption-line spectra are represented by open squares
and emission-line galaxies are shown as crosses. The
open circle represent the galaxy whose spectrum could not
be inspected. As for A3565, there are very few galaxies with
emission lines in A3560, most of them being found in the outer
regions of the cluster.
}
\end{figure}

\begin{figure}
\plotone{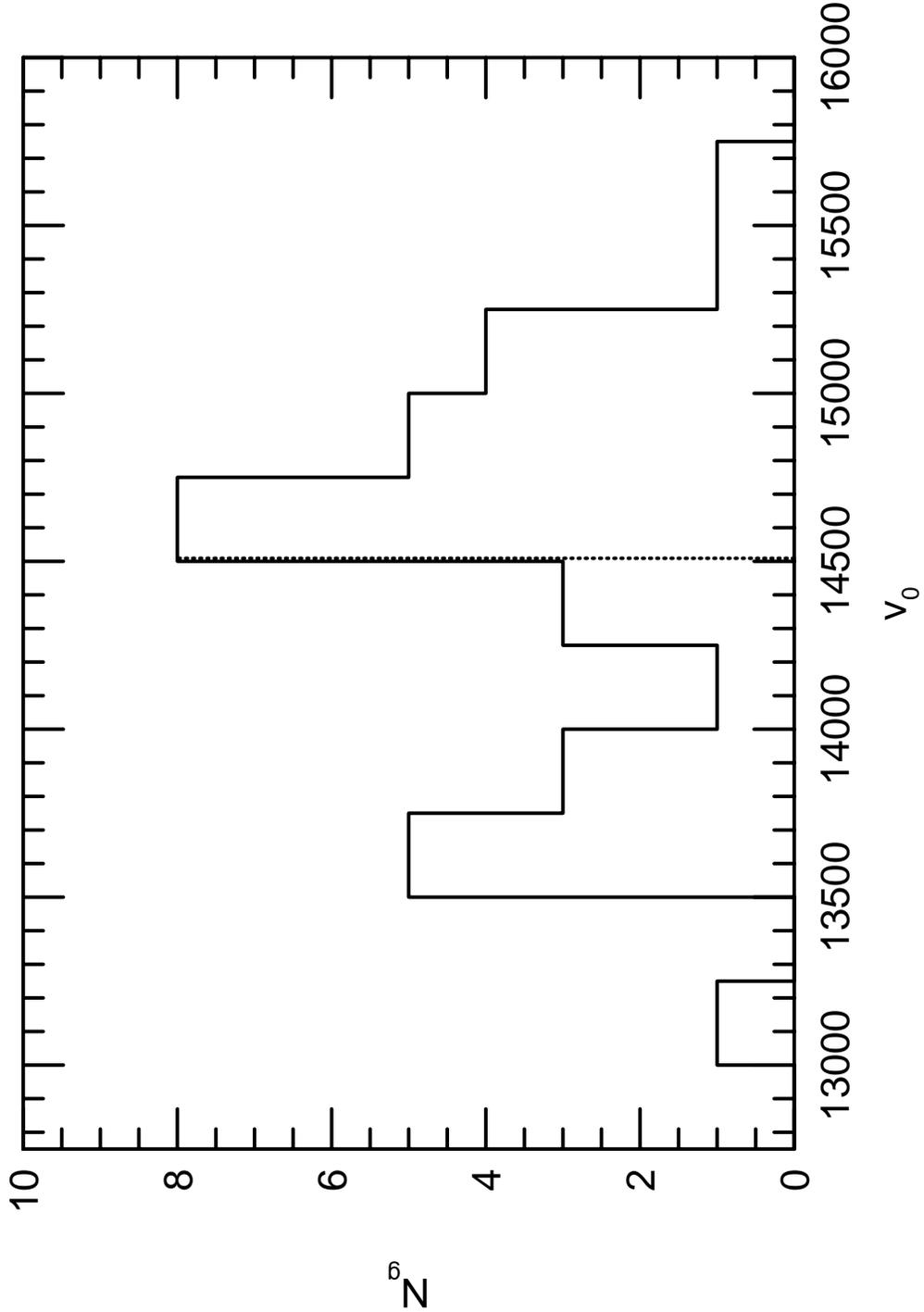}
\caption{
Radial Velocity distribution of galaxies in the
region 13$^h 28^m \leq \alpha \leq  13^h 33^m$ and
-33\deg 30$' \leq \delta \leq$  -32\deg 20$'$, in
250 \kms bins. The dotted line shows the radial velocity of
the dumbbell galaxy formed by objects WMMA 032 and WMMA 033 in Table 4.
}
\end{figure}

\begin{figure}
\plotone{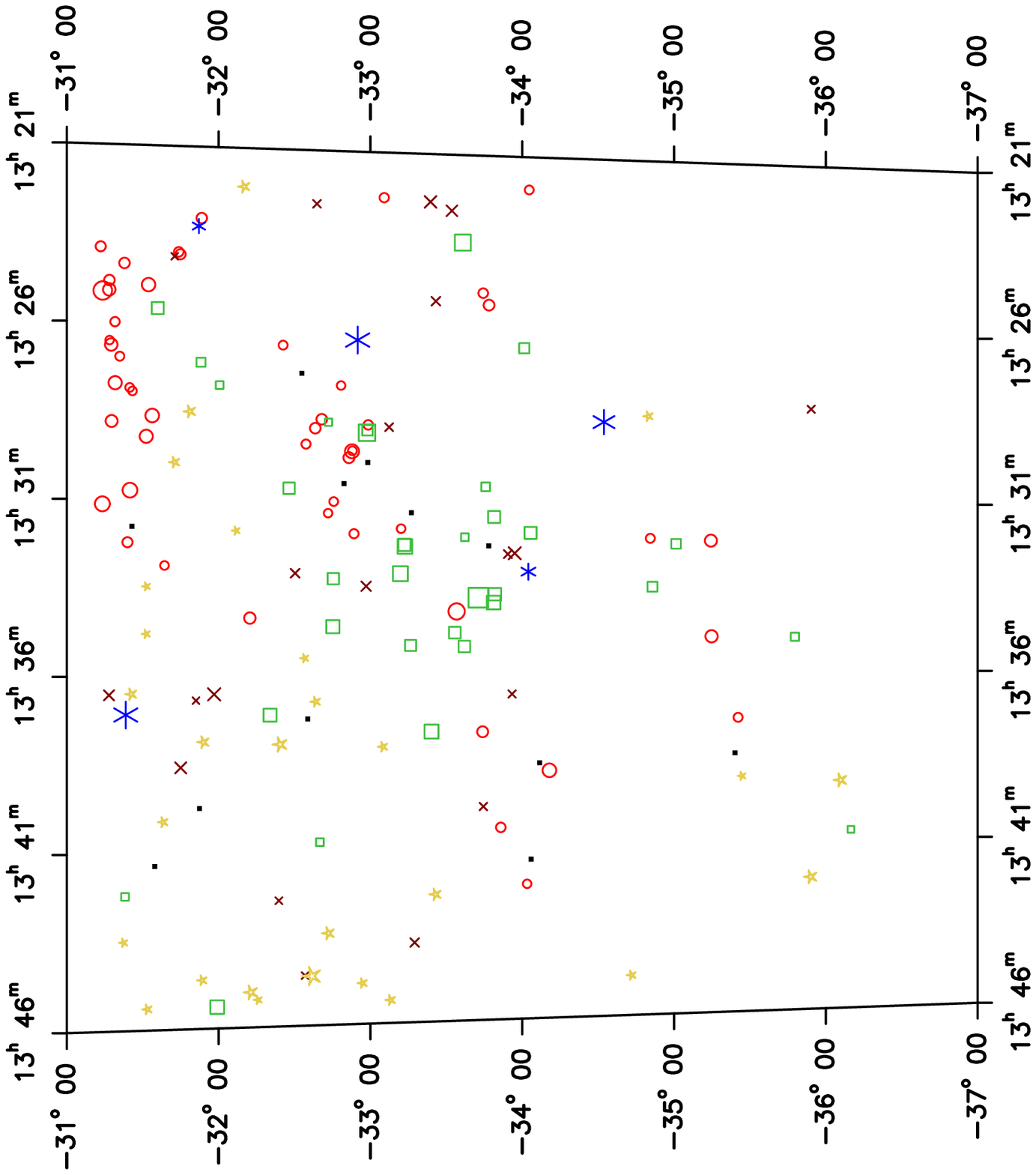}
\caption{
Distribution of galaxies brighter than $m_B$ = 15.5 $mag$ in a 5\deg
$\times$ 5\deg region centered on IC~4296. The symbols have the same
meaning as in Figure 2. The figure shown no significant increase in
the number of galaxies in the redshift interval of 3000 to 5000 \kms
(open squares) which could belong to A3565.
}
\end{figure}


\begin{references}
\normalsize

\reference {} Abell, G. O., Corwin, H.G., \& Olowin, R.P.  1989, ApJS,
70, 1

\reference {} Alonso, M. V., da Costa, L. N., Pellegrini, P. S. \& Kurtz, M. J.
1993, AJ 106, 676

\reference {}  Arp H. C., \& Madore, B. F.  1987, ``A Catalogue of Southern
Peculiar Galaxies and Associations'', (Cambridge: Cambridge Univ. Press) (AM)

\reference {} Assafin, M., , Martins, R. Vieira, \& Andrei, A. H. 1997,
AJ, 113, 1451

\reference {} Bardelli, S., Pisani, A., Ramella, M., Zucca, E., \&
Zamorani, G. 1998, MNRAS 300, 589

\reference {} Beers, T. C., Flynn, K. \& Gebhardt, K. 1990, AJ, 100, 32


\reference {} Bothun, G., Geller, M. J., Beers, T. C. \& Huchra,
J. P. 1983, ApJ 268, 47

\reference {} Brown, D., L., \& Burns, J. O. 1991, AJ, 102, 1917

\reference {} Buote, D. A., \& Fabian, A. C.,  1998, MNRAS, 296, 977

\reference {} Buta, R. 1995,  ApJS, 96, 39 (CSRG)

\reference {} Corbin, T. \& Urban, S. 1988, in I.A.U. Symp. No. 133,
``Mapping the Sky'', ed. H. Eichhorn (Dordrecht: Reidel), 75

\reference {} Cot\'e, S., Freeman, K. C., Carignan, C., \& Quinn,
P. J.  1997, AJ, 114, 1313


\reference {} da Costa, L.N., Pellegrini, P.S., Nunes, M.A., Willmer,
C., Chincarini, G., \& Cowan, J.  1986, AJ, 91, 6 (DC1)

\reference {} da Costa, L.N., Pellegrini, P. S., Willmer, C., de Carvalho, R.,
Maia, M., Latham, D. W. \& Geary, J. C. 1989, \aj, 97, 315

\reference {} da Costa, L.N., Willmer, C. N. A., Pellegrini, P.S., 
Chaves, O. L., Rit\'e, C., Maia, M. A. G., Geller, M. J., Latham, D. W.,
Kurtz, M. J., Huchra, J. P., Ramella, M., Fairall, A. P., Smith, C., \&
L\'\i pari, S.   1998, AJ, 116, 1

\reference {} da Costa, L.N., Willmer, C., Pellegrini, P.S., \&
Chincarini, G.  1987, AJ, 93, 1338 (DC2)

\reference {} David, L. P., \& Blumenthal, G. R.  1992, ApJ 389, 510

\reference {} Danese, L., De Zotti, G., \& di Tullio, G.  1980, A\&A,
82, 322

\reference {} Dressler, A., \& Shectman, S. A.  1988, AJ, 95, 985

\reference {} Drinkwater, M. J., Proust, D., Parker, Q. A., Quintana,
H., \& Slezak, E. 1999, PASAu, in press; astro-ph/9903028

\reference {} Ebeling, H., Voges, W., B\"ohringer, H., Edge, A. C.,
Huchra, J. P. \&  Briel, U. G. 1996, MNRAS, 281, 799

\reference {} Ettori, S., Fabian, A. C., \& White, D. A.  1997, MNRAS,
289, 787

\reference {} Forman, W., Jones, C., \& Tucker, W.  1985, ApJ, 293, 102

\reference {} Garcia, A. M.  1993, A\&AS, 100, 47


\reference {} Graham, A. W., Lauer, T. D., Colless, M. \& Postman,
M. 1996, ApJ 465, 534

\reference {} Gregorini, L.,  de Ruiter, H. R., Parma, P., Sadler,
E. M., Vettolani, G., \& Ekers, R. D.  1994, A\&AS, 106, 1

\reference {} Heisler, J., Tremaine, S., \& Bahcall, J. N.  1985, ApJ,
298, 8


\reference {} Huchra, J. P., \& Geller, M. G.  1982, ApJ, 257, 243

\reference {} Jarvis, J.F., \& Tyson, J.A.  1981, \aj, 86, 476

\reference {} J\o rgensen, I., Franx, M., \& Kjaergaard, P.  1995,
MNRAS 276, 1341

\reference {} Kemp, S. N.  1994, A\&A, 282, 425

\reference {} Kemp, S. N., \&  Meaburn, J.  1991, MNRAS, 252, 27

\reference {} Kemp, S. N., \&  Meaburn, J.  1993, A\&A, 274, 19

\reference {} Killeen, N. E. B., \& Bicknell, G. V.  1988, ApJ, 324, 198

\reference {} Koranyi, D. M., Geller, M. J., Mohr, J. J., \& Wegner,
G.  1998, AJ, 116, 2108

\reference {} Kurtz, M. J. \&  Mink, D. J. 1998, PASP, 110, 934

\reference {} Kurtz, M. J., Mink, D. J., Wyatt, W. F., Fabricant, D. G.,
Torres, G., Kriss, G. A., \& Tonry, J. L. 1992, in ASP Conf. Ser. Vol. 25,
Proc. 1$^{st}$ Ann. Conf. Astronomical Data Analysis Software and
Systems, ed., D.M. Worral, C. Biemesderfer, \& J. Barnes
(San Francisco: ASP), 432

\reference {} Lasker, B.M., Sturch, C.R., McLean, B.M., Russel, J.L.,
Jenker, H., \& Shara, M. 1990, \aj, 99, 2019 (HSTGC)

\reference {} Lauberts, A., \& Valentijn, E.A.  1989, The Surface
Photometry Catalogue of the ESO-Uppsala Galaxies, (Garching: ESO) (ESO--LV)

\reference {} Lauer, T. R. \& Postman, M. 1994, ApJ, 425, L418 

\reference {} Lauer, T. R., Tonry, J. L., Postman, M., Ajhar, E. A. \&
Holtzman, J. A. 1998, ApJ 499, 577

\reference {} Lin, H., Kirshner, R. P., Schectman, S. A., Landy, S. D.,
Oemler, A., Tucker, D. L. \& Schechter, P. L. 1996, ApJ, 471, 617

\reference {} Lynden-Bell, D., Faber, S.M., Burstein, D., Davies,
R.L., Dressler, A., Terlevich, R.J., \& Wegner, G.  1988, ApJ, 326, 19

\reference {} Lynden-Bell, D., Lahav, O., \& Burstein, D.  1989,
MNRAS, 241, 325

\reference {} Maia, M.A.G., da Costa, L.N., \& Latham, D.W.  1989,
ApJS, 69, 809

\reference {} Massey, P. 1992, A User's Guide to CCD Reductions with
IRAF (Tucson: KPNO Computer Support Group)

\reference {} Maddox, S. J., Efstathiou, G. \& Sutherland, W. J. 1990,
MNRAS 246, 433

\reference {} Mahdavi, A., Geller, M. J., Boehringer, H., Kurtz, M. J. \& 
Ramella, M. 1999, astro-ph/9901095

\reference {} Marston, A. P.  1988, MNRAS, 230, 97

\reference {} Matthewson, D. S., Ford, V. L. \& Buckhorn, M.  1992,
ApJS, 81, 413

\reference {} Melnick, J.  \& Moles, M.  1987, RMxAA, 14, 72

\reference {} Metcalfe, N., Godwin, J. G., \& Peach, J. V. 1994,
MNRAS, 267, 431

\reference {} Mills, B. Y., Slee, O. B., Hill, E. R.  1960,
Aust.J. Phys. 13, 676

\reference {} Mount Stromlo Abell Cluster Supernova Search Team, 1997,
IAU Circular 6708A

\reference {} Mould, J. R., Staveley-Smith, L., Schommer, R. A.,
Bothun, G. D., Hall, P. J., Han, M. S., Huchra, J. P., Roth, J.,
Walsh, W., \& Wright, A. E.  1991, ApJ, 383, 467

\reference {} Mulchaey, J. S., Davis, D. S., Mushotzky, R. F., \&
Burstein, D.  1996, ApJ, 456, 80

\reference {} Mulchaey J. S., \& Zabludoff, A., I.  1998a, ApJ, 496, 73

\reference {} Mulchaey J. S., \& Zabludoff, A., I.  1998b, astro-ph/9810458

\reference {} Nolthenius, R. 1993, ApJS, 85, 1

\reference {} Paturel, G., Petit, C., Kogoshvili, N., Dubois, P., Bottinelli,
 L., Fouque, P., Garnier, R. \& Gouguenheim, P. 1991, A\&AS, 91, 371

\reference {} Pierre, M., B\"ohringer, H., Ebeling, W., Vosges, W.,
Schuecker, P., Crudacce, R., \& MacGillivray, H.  1994, A\&A, 290, 725

\reference {} Pinkney, J., Roettinger, K, Burns, J. O., \& Bird,
C. M.  1996, ApJS 104, 1

\reference {} Postman, M. \& Lauer, T. R., 1995, ApJ, 440, 28 

\reference {} Quintana, H.,  Ramirez, A., Melnick, J.,  Raychaudhury,
S., \& Slezak, E. 1995 AJ 110, 463 (QRM)

\reference {} Ramella, M., et al. 1999, A\&A, 342, 1

\reference {} Ramella, M., Geller, M. J., \& Huchra, J. P.  1989, ApJ,
344, 57

\reference {} Richter, O. -G.  1984, A\&AS, 58, 131

\reference {} Richter, O. -G.  1987, A\&AS, 67, 261


\reference {} R\"oser, S., \& Bastian, U.  1991, PPM Star Catalogue
(Heidelberg: Spektrum Akademischer Verlag)

\reference {} Saglia, R. P., Bertin, G., Bertola, F. Danziger, J.,
Dejonghe, H., Sadler, E. M., Stiavelli, M., de Zeeuw, P. T.,
Zeilinger, W. W.  1993, ApJ 403, 567

\reference {} Sandage, A.  1972, ApJ, 178, 1



\reference {} Schlegel, D. J., Finkbeiner, D. P. \& Davis, M., 1998, ApJ,
500, 525

\reference {} Theureau, 1998, A\&AS, 130, 333 (T98)



\reference {} Valdes, F. 1982, Focas User's Manual (2$^{nd}$ ed.;
Tucson: KPNO Computer Support Group)
 
\reference {} Valentijn, E. A., \& Casertano, S.  1988, A\&A 206, 27


\reference {} Vettolani, G., Chincarini, G., Scaramella, R. \&
Zamorani, G.  1990, AJ, 99, 1709

\reference {} Wegner, G. et al. 1999, {\it{in preparation}}.

\reference {} West, M. J., Oemler, A., \& Dekel, A.  1988, ApJ, 327, 1

\reference {} Willmer, C.N.A., Focardi, P., Chan, R., Pellegrini,
P.S., \& da Costa, L.N.  1991, AJ 101, 57

\reference {} Yahil, A., Vidal, N. V.  1977, ApJ, 214, 347

\reference {} Zabludoff, A., I., \& Mulchaey, J., S.  1998a, ApJ, 496, 39

\reference {} Zabludoff, A., I., \& Mulchaey, J., S.  1998b, ApJ, 498,
 5

\end{references}
\end{document}